\documentclass[lettersize,journal]{IEEEtran}
\usepackage{amsmath,amsfonts}
\usepackage{algorithmic}
\usepackage{algorithm}
\usepackage{array}
\usepackage[caption=false,font=normalsize,labelfont=sf,textfont=sf]{subfig}
\usepackage{textcomp}
\usepackage{stfloats}
\usepackage{url}
\usepackage{verbatim}
\usepackage{graphicx}
\usepackage{cite}

\usepackage{paralist, tabularx}
\usepackage{amsmath}
\usepackage{amsthm}

\DeclareMathOperator*{\argmin}{arg\,min}

\begin{document}

\title{(Im)balance in the Representation of News? An Extensive Study on a Decade Long Dataset from India}

\author{\IEEEauthorblockN{\textbf{Souvic Chakraborty}}
    \IEEEauthorblockA{chakra.souvic@gmail.com}\\
    \and
    \IEEEauthorblockN{\textbf{Pawan Goyal}}
    \IEEEauthorblockA{pawang.iitk@gmail.com}\\
    \and
    \IEEEauthorblockN{\textbf{Animesh Mukherjee}}
    \IEEEauthorblockA{animeshm@gmail.com\\
    \textit{Department of Computer Science \& Engineering} \\
    \textit{Indian Institute of Technology,}
    Kharagpur, India \\
    }}

\markboth{Journal of \LaTeX\ Class Files,~Vol.~14, No.~8, August~2021}%
{Shell \MakeLowercase{\textit{et al.}}: A Sample Article Using IEEEtran.cls for IEEE Journals}


\maketitle

\begin{abstract}
(Im)balance in the representation of news has always been a topic of debate in political circles. 

The concept of balance has often been discussed and studied in the context of the social responsibility theory and the \textit{prestige press} in the USA. While various qualitative as well as quantitative measures of balance have been suggested in the literature, a \textit{comprehensive analysis} of all these measures across a \textit{large dataset} of the \textit{post-truth era} comprising \textit{different popular news media houses} and \textit{over a sufficiently long temporal scale} in a non-US democratic setting is lacking. We use this concept of balance to measure and understand the evolution of imbalance in Indian media on various journalistic metrics on a month by month basis. For this study, we amass a huge dataset of over \textit{four million} political articles from India for 9+ years and analyze the extent and quality of coverage given to issues and political parties in the context of contemporary influential events for three leading newspapers.  We use several state-of-the-art NLP tools to effectively understand political polarization (if any) manifesting in these articles over time. We find that two out of the three news outlets are more strongly clustered in their imbalance metrics. We also observe that only a few locations are extensively covered across all the news outlets and the situation is only slightly getting better for one of the three news outlets. Cloze tests show that the changing landscape of events get reflected in all the news outlets with border and terrorism issues dominating in around 2010 while economic aspects like unemployment, GST, demonetization, etc. became more dominant in the period 2014 -- 2018. Further, cloze tests clearly portray the changing popularity profile of the political parties over time.

\end{abstract}

\begin{IEEEkeywords}
Media Bias, India, Politics
\end{IEEEkeywords}

%


\maketitle
\section{Introduction}

Indian media has been the target of blatant criticism internationally for propagation of one-sided views on specific issues and deliberate introduction of imbalance and sensationalism in reporting news\footnote{\url{https://www.aljazeera.com/opinions/2020/2/24/indias-media-is-failing-in-its-democratic-duty}}. Specific media houses have been consistent targets by the supporters of specific parties for propagating unbalanced views on issues benefiting some specific party. Many of these unverified news and systematic imbalances have been criticized to introduce communal disharmony and even loss of lives.

So, we intend to study how (im)balanced Indian news media is and how it has changed its course over time with different ruling parties in the centre and on face of large scale events including the national election in 2014 with almost a billion voters and significant change in share of seats for parties, large scale changes in monetary policy like demonetisation and introduction of GST in 2016 \& 2017 and nation-wide non-political anti-corruption movement in 2011 with huge influence and mass following having significant impact on later political discourse.

While popularity of sharebaits is a problem, a bigger problem is perhaps the reliance of the common mass on social media to get their daily news feeds~\cite{foot1}. There is an increasing lack of diversity in the algorithmically driven news feeds that the social media typically presents to its readers. In many cases this leads to reinforcing the `bias' among readers in the form of increased polarity of the political opinion of these readers over time~\cite{bakshy15}. This idea has been presented in the past literature in various forms like echo chambers~\cite{garrett09,flaxman16} and filter bubbles~\cite{resnick13}. Had the nature or inclination of bias used by newspapers been dynamic temporally or topically, it would have been less likely to influence people at large. However, recent research~\cite{vosoughi2018spread} in bias propagation suggests that typically ``static'' forms of biased news are more likely to spread. So, once a specific bias is introduced by the major media houses systematically, it will act as a positive feedback loop catering to the ``confirmation'' of their readers and making it more difficult for the media houses to introduce a different stance contradicting the view of their readers which they themselves shaped and are identified with. 

\subsection{Current studies and their limitations}
Most of the above investigations relating to echo chambers, filter bubbles and confirmation bias deal with the end effect of introducing systematic bias in various forms. A crucial question is how this bias is introduced in the first place? Can the source of this bias be suitably quantified? We posit, in this paper, that the `imbalance' in the representation of the news could be a potential source for this bias. This is motivated by the works of previous researchers on fairness in journalism who often consider bias as the opposite of accuracy, balance, and fairness ~\cite{fico1994,fico1995,lacy91,simon1989,Streckfuss1990}. 

While `imbalance' is a useful measure for quantification of such bias as demonstrated in the works of previous researchers on fairness in journalism who often consider bias as the opposite of accuracy, balance, and fairness ~\cite{fico1994,fico1995,lacy91,simon1989,Streckfuss1990}, imbalance can have many facets and must be examined from different angles unlike many influential works~\cite{lacy91,d2000media} which examined news articles on variations of coverage bias.

Computational studies (apart from the manual studies done by independent journalists) in news media~\cite{Budak2014,Spillane17,lazer2018science} done till now are at most times lacking in either the number of articles examined or the time span of examination or both. One of the prime limitations of these studies is that most of them have taken a piecemeal approach. In order to overcome that limitation, we examine our corpus under various lenses of bias with variety of tools in order to get a more nuanced view of the evolution of Indian media and society at large over the last decade. 



\subsection{Objectives of this work}

The primary objective of this paper \textit{is to examine a large body of news articles through various lenses of biases.} While most of the news media datasets available are specific to US~\cite{us1, us2, us3} or Europe~\cite{eu16,eu2} or Austraila~\cite{aus18}, they are limited to the specific regional scope in time/events, diversity of media groups etc. So, for this specific study, we mine Indian newspapers with highest readership numbers and online availability that spans over almost a decade. We make efforts to understand the temporal behavior of different forms of bias featuring across three news media outlets by analyzing this massive news dataset, in the Indian context. We summarize our contributions in the following section.

\subsection{Key contributions}
In the following we summarize our specific contributions.

\begin{compactitem}

    \item We collect news articles for three leading Indian newspapers for a huge span of 9+ years. In addition, for every article we also separately collect various metadata like the source of the article, the headline and the URL. 
    \item We use seven different metrics of (im)balance that are easy to compute as well as interpret, motivated by \textit{two completely orthogonal viewpoints} generally upheld by researchers or termed important by media bias gatekeeping organizations. 
    \item We compute these metrics for (im)balance month-wise for the three leading newspapers and observe the temporal trends for each kind of metrics, cluster these time series and discuss the most interesting observations.
    \item We use different NLP tools like word embedding association test and masked language modelling to answer several research questions in an India specific setting and discuss the implications of the results obtained. These investigations unfold various interesting trends in Indian political discussions over the years.

 
\end{compactitem}%

\if{0}Our contribution in this paper can be summarized in 3 points:

1. Collection of large dataset with content, headline, source and number of link-shares on twitter.

2. Performing three-way annotation for transfer learning on the Hyperpartisan bias detection task 

3. Performing large scale correlation study and establishing the 

Finally, we hope that our study will help our fellow researchers to understand the efficacy of the peer-reviewing process and will inspire people to think of ways to improve the same.

\section{Related Works}
Researchers have been studying the perceived bias and credibility of the mass media for a long time ~\cite{jacobson1969mass,erskine1970polls,grossberg2006mediamaking,ladd2012americans,kalbfleisch2003credibility,gronke2007disdaining,earl2001assessing}. 
In fact, bias is a central topic of discussion in the field of journalism and communication~\cite{Yano:2010:SLB:1866696.1866719,entman2007framing,scheufele1999framing}. Whereas studies on bias and the idea of journalistic truth in news media can be found as early as 1922~\cite{lippmann1922public}, systematic studies to model the possible influencers of bias in media and the effect of the bias on society can be found in ~\citet{herman1988manufacturing}. In their famous work,~\citet{herman1988manufacturing} introduced the propaganda model for the manufacture of public consent. They formulated and described five editorially distorting filters to model the distortions and bias introduced by the media house to the actual news. Despite criticisms\cite{f5} from mainstream media, their work seem to remain relevant even in the digital age~\cite{article2009}.
However, these works do not deal with the need for computational efficiency and automation in dealing with bias. While many watchdog groups\cite{f6}\cite{f7} have been established by eminent journalists throughout the globe, they can also be accused of bias as the study involves human beings and the editors themselves may have subtle or explicit biases. 

The main obstacle toward taking a computational approach in defining bias may be the very fundamental subjectivity in a linguistically abstract word like ``bias''. While most people concur on characterization of ``media bias''  using correlated terms and definitions, there have been very different studies on the empirical definition of the same~\cite{eu2,Saez_Trumper,us5,cov6,ribeiro2018media}. People from the media have been largely against definite characterizations right from criticisms stemming against the work of~\citet{herman1988manufacturing} to the Guardian's recent claim, ``Media bias is ok if honest''\footnote{https://www.theguardian.com/commentisfree/2019/sep/10/media-bias-is-ok-if-its-honest}. Here the journalists from the Guardian point out the inherent bias in reporting a specific story as it is difficult to decide if reporting a specific story in all its importance should be called bias or reporting every story (even the trifle ones) with equal coverage should be called bias. 

There have been numerous studies examining bias in media especially in the US and the European context. While the term ``bias'' still remains abstract, some studies have put efforts to make a distinction between the computational sense of bias and the journalistic sense of the same making it more scientifically definable and quantifiable.
Journalistic and linguistic studies mostly discuss selection/coverage bias, confirmation/statement bias~\cite{lazaridou2017identifying,lin2011more,nickerson1998confirmation,Saez_Trumper} and psychological/cognitive
biases~\cite{caliskan2017semantics,recasens2013linguistic}. Recently, a lot of works are being done where the researchers are interested to formulate a computational basis for investigating bias. Some works are focused on specific kinds of bias,
such as gender~\cite{Bolukbasi,madaan2018analyze,zhao2017men} and race~\cite{chouldechova2017fair}. Politics, in particular, is a widely studied and discussed topic. Researchers seek to find ideological political bias of users in social networks~\cite{conover2011predicting,johnson2016identifying,wong2016quantifying}, news media~\cite{baly2018predicting,budak2016fair,laver2003extracting,le2017scalable,ribeiro2018media} and user comments~\cite{ribeiro2017everything,Yigit_Sert}.
D'Alessio and Allen~\cite{d2000media} list three kinds of media bias to be the most widely studied. Coverage/Visibility bias~\cite{3Jakob}, gatekeeping bias/selectivity~\cite{6Richard} or selection bias~\cite{7Groeling} (sometimes referred to as agenda bias~\cite{3Jakob}) and statement bias/tonality bias/presentation bias~\cite{3Jakob,7Groeling}.

Of late the CSCW community has also started conducting active research in this area. In~\cite{Morgan2013}, the authors study how users share news articles on Twitter. They observe that while the users are biased in consuming only specific types of news they are not biased in sharing news items; in fact, there is a large diversity in whatever news items they share on Twitter. In~\cite{Grevet:2014} the authors study how political differences in social media can be managed. They study the political disagreements on Facebook to identify situations when diverse political opinions can effectively coexist in the online world. In~\cite{Kulshrestha2017} the authors study the sources of bias in political searches in social media. They specifically attempt to distinguish the bias in the ranking of the search results that are caused by the input data and the ranking algorithm. In~\cite{Robert2018} the authors study partisan audience bias in Google search patterns. The authors report that they observe the well-known ``filter bubble''\cite{f9} phenomenon in their results.  

We attempt to formulate definitions of ``bias'' compiling relevant ideas from this huge corpus of past research as well as adding necessary new formulations and apply these on the large scale longitudinal Indian news media dataset that we have collected. This analysis allows us to obtain various interesting insights about the extent of different forms of bias present across widely popular Indian media houses.
\fi
\section{Background}\label{related_works}

In this section we first present a brief review of fairness and balance in the journalism literature. This is followed by a narration of how we build up our work on these ideas.

\noindent\textbf{(Im)balance in journalism}: Fairness and balance in the press has been studied for a long time in the journalism literature. One of the early works pertains to how the prestige press is distinctly different from the media outlets with wide circulation~\cite{lacy91}. The authors perform a small scale data-driven study to show that the prestige press presents a more balanced coverage of local stories compared to wide circulation media outlets. Fico et al~\cite{fico1994} studied the newspaper coverage of US in Gulf war and find that wide circulation newspapers were more likely to favor anti-war advocates than smaller ones. Fico et al~\cite{fico1995} develop a content based technique to study the newspaper coverage of controversial issues. They find that only 7\% of the stories were evenly balanced and the coverage on the Gulf war issue was most imbalanced. Fico et al~\cite{fico1999} study structural characteristics of newspaper stories on the 1996 US presidential election. One of the very important findings in this work is that event coverage was the biggest predictor of imbalanced story. Fico et al~\cite{fico2008} study balance in election news coverage about 2006 US senate elections. They observe that women reporters provided more evenly balanced treatment of the candidate assertions. Carter et al~\cite{carter2002} the authors report the structural balance in local television election coverage. Geri et al~\cite{geri2008} the authors study the broadcast and cable network news coverage of the 2004 presidential election and find that broadcast networks were more balanced in their aggregate attention to the presidential candidates compared to the cable networks. In \cite{Robert2018}, the researchers studied Google searches to find partisan bias in widely used digital platforms. Morgan et al\cite{Morgan2013} studied bias in social media shares and of news items.On the other hand, Kulshrestha et al \cite{Kulshrestha2017} studied social media bias for political searches.
Finally, in ~\cite{resnick13} the authors study remedy techniques to limit exposure to bias : \textit{strategies for promoting diverse exposure}.

\noindent\textit{The present work}: In most of the above works, (im)balance has largely been quantified in terms of coverage. However, we posit that (im)balance can manifest in many different ways. In the following we outline these assimilating concepts from various past literature. 

\noindent\textit{Coverage imbalance}: Coverage imbalance is the extent to which some specific entities/topics are covered in the articles by a specific media house. Coverage imbalance may originate from the inequality in the number of articles published with stories related to each major party or the amount of inkspace given to each party (even if the number of articles are equal, articles covering one party can be longer) or the amount of inkspace used to cover speeches of leaders of each political party.
  
  For the first two metric of coverage imbalance, we take inspiration from the work of D'Alessio et al ~\cite{d2000media} and Lacy et al ~\cite{lacy91}. The equivalent of number of stories featuring a party is the number of headlines featuring that party, the ideal contender of measuring Gatekeeping bias~\cite{d2000media}. On the other hand, we take the sentences in the content as proxy for a combined measure of fairness and balance in work of Lacy et al~\cite{lacy91} to measure the amount of inkspace given to a particular party. For all the analysis we use words as the analog of inkspace here as words are the basic units of semantic information in online media.

 Following the work of Lin et al\cite{lin-etal-2006-side}, we introduce another measure of coverage imbalance, the \textit{point of view} imbalance. The point of view from which a news story is reported matters as the editors have to choose selective viewpoints due to the constraint in space/number of words that can be used in a readably long article. While one newspaper may choose to quote the government sources, the same news story may be reported quoting the opponent leader. Thus, imbalance gets introduced if diverse viewpoints are not equally represented as discussed in the guidelines for fair reporting by FAIR\footnote{\url{https://www.fair.org}}. We try to capture the sentences where speech of any person affiliated to a party is reported and calculate the number of words (semantic equivalent of inkspace) just like the previous cases of coverage imbalance.

\noindent\textit{Tonality imbalance}: Choice of words matters in presentation of a news. Same piece of news can be presented with positive sentiments expressed toward a specific entity; likewise it can be presented in a way that would portray the specific entity in a negative light. The views of the editors are often politically biased as evident from the ``opinion'' columns of the newspapers and can get reflected in the news story they are covering also. So, it is important to check for tonal imbalance in the news-stories.

Different distributions of sentiments for different topics are often used~\cite{biascred} to detect bias and credibility of news sources. We can use similar measures to capture tonality imbalance. We do so in two possible ways here: by measuring the density of \textit{positive \& negative sentiments} in the text or by measuring the overall density of \textit{subjectivity} of the text. We use average sentiment/subjectivity over all sentences, talking about any specific party, weighted by the number of words here as a quality of a long sentence should have proportionate presence in the metric of our choice to express the actual inkspace equivalency.

In addition to the introduction of different perspectives on imbalance as above, we also upscale our study in two other major directions. First, we perform the analysis on a huge dataset comprising three Indian news outlets leading to a total of 3.86 million articles. Second, rather than aggregate statistics, we present temporal characteristics of the imbalance which allows us to make various important and nuanced observations.  

\section{Dataset}\label{data}
From the list of top news media by readership, published by the Indian Readership Survey (IRS) 2017~\cite{f10} compiled by the Media Research Users Council (MRUC), we collect the news articles data for three popular English language newspapers in India, namely, Times of India (TOI)\footnote{https://timesofindia.indiatimes.com/archive.cms}, The Hindu\footnote{https://www.thehindu.com/archive/} and India Today\footnote{https://www.indiatoday.in/archives/}, where an online archive is available. We create a date-wise repository of more than four million news articles spanning 9$+$ years (2010--2018) of news data crawling through the archives.

The total number of articles retrieved for TOI, The Hindu and India Today are 1,899,745, 1,032,377 \&  926,922, respectively. A brief description of the year wise statistics of the political articles across the three newspapers is noted in Table~\ref{yearlystats}.

\begin{table}[h]
\centering
\small
\begin{tabular}{l|l|l|l}
\hline
 & TOI  & The Hindu & India Today \\ \hline
 2010 & 101903 & 74927 & 45297 \\ \hline
 2011 & 169183 & 75160 & 27135 \\ \hline 
 2012 & 236548 & 86107 & 146634 \\ \hline
 2013 & 237372 & 88795 & 197139 \\ \hline
 2014 & 210859 & 106013 & 152260 \\ \hline
 2015 & 243473 & 204421 & 135898 \\ \hline
 2016 & 208861 & 192120 & 71155 \\ \hline
 2017 & 242334 & 117584 & 76800 \\ \hline
 2018 & 249212 & 87250 & 74604 \\ \hline
\end{tabular}%

\caption{\label{yearlystats}Year-wise statistics of the collected data for the chosen time range for all three newspapers. Note that for India Today the pattern is unique in that from 2011 to 2012 the circulation became 5x while from 2015 to 2016 the circulation became 0.5x. None of the other outlets indicate such stark shifts.} 
\end{table}

\subsection{Pre-processing of the raw dataset} Our data consists of the headlines of the news stories, date of publication and the content. 

\noindent\textit{Keyword based shortlisting of articles}: Our entire study is done to depict the political imbalance of different media houses in representation of news related to two major political parties of India -- `Bharatiya Janata Party'\footnote{\url{https://en.wikipedia.org/wiki/Bharatiya_Janata_Party}} and `Indian National Congress'\footnote{\url{https://en.wikipedia.org/wiki/Indian_National_Congress}}, more famous as `BJP' \& `Congress'. 
We present a keyword based analysis throughout all the metrics. The set of keywords chosen for BJP (is referred to as BJP\textsubscript{keywords} from now on) consists of --
`Bharatiya Janata Party', `BJP', `Akhil Bharatiya Vidyarthi Parishad', `ABVP', `National Democratic Alliance', `NDA' .
The set of keywords chosen for Congress (is referred to as Congress\textsubscript{keywords} from now on) consists of -- `Congress' (the most popular version of the full name of the party), `INC', `National Students' Union of India', `NSUI', `United Progressive Alliance' and `UPA'.


\noindent\textit{Motivation for the choice of keywords}: A natural question would be why we chose only the above keywords for our experiments. The choice is motivated by the additional set of experiments that we did to fix the above set of keywords. We first chose these keywords on the basis of two criteria. We manually checked the news articles to find the important indicators and through this qualitative analysis, we found that the discussion around the parties can typically be identified with the most popular version of their names including acronyms \& full names and popular acronyms of coalition governments \& student organizations.  Next we included the names of the personalities related to the party thereafter and observed the results are completely swayed away by the coverage of influential figures like the names of the Prime Minister or all the other names holding important offices. We argue that the political parties are distinct from the personalities affiliated to those parties or the government formed by those parties. This justifies the exclusion of these entities from our seed set since they might correspond to issues related to (i) the functioning of the government and not the party in particular or (ii) the functioning of the personalities who might hold different portfolios within the government and might have their own charismatic presence on many issues and may have a different face than being a party member only while commenting on different issues. Thus we argue that we should limit the keywords to the most popular seed set chosen alone that includes party names, acronyms, names of student wings and name of the democratic alliances where any of the two parties are the most influential ones.

\section{Methodology}
This section is laid out in three parts. In the first part we discuss the different metrics of imbalance. In the second part, we demonstrate how the different temporally varying metrics of imbalance can be summarised to reflect certain universal patterns.  Finally, we outline a method to identify word associations that could reflect polarisation.   
\subsection{Uniform metric of imbalance}
We adopt the method of determining imbalance from the acclaimed work of Lacy et al\cite{lacy91}. For each of the metric $i$, that we use in the subsequent analysis, (henceforth, $metric_i$), we compute an imbalance score at the granularity of each month. Thus, for a particular month of a year, the imbalance score (henceforth, $imbalance\textsuperscript{i}\textsubscript{yy-mm}$) is calculated as follows, adapting the work of Lacy et al\cite{lacy91}.

\begin{align}
Imbalance\textsuperscript{i}\textsubscript{yy-mm} &= \frac{Score\textsubscript{i}(B\textsuperscript{i}\textsubscript{yy-mm})-Score\textsubscript{i}(C\textsuperscript{i}\textsubscript{yy-mm})}{Score\textsubscript{i}(B\textsuperscript{i}\textsubscript{yy-mm})+Score\textsubscript{i}(C\textsuperscript{i}\textsubscript{yy-mm})}\label{e1}
\end{align}

where $B\textsuperscript{i}\textsubscript{yy-mm}$ and $C\textsuperscript{i}\textsubscript{yy-mm}$ correspond to the \textit{aggregated documents} for the two political parties considered (BJP and Congress, respectively, in our case). We detail a criterion to include the sentences of the contents/headlines in any/both of the two documents for each of the metrics.
We also detail a criterion to determine the score for that document for each of the metric, which helps us compute the two scores in the equation above. 

Next, we obtain the imbalance scores for each metric across the timeline of 2010--2018. Note that this \textit{imbalance score has a direction}. All \textit{positive scores correspond to a leaning toward BJP} and \textit{all the negative scores correspond to a leaning toward Congress}. The \textit{absolute value of the score} denotes the \textit{imbalance without direction}. Apart from the timeline plots illustrating the imbalance with direction, we also compute the aggregate values of the absolute imbalance score for each metric averaged over the timeline of computation. 

\subsection{Coverage imbalance} We have done two studies to find balance in coverage of political parties in newspapers, one on the basis of (i) the \textit{headlines} and the other on the basis of (ii) the \textit{content} of the article. 

\noindent\textbf{Headlines} 

\noindent\textit{Sentence inclusion criterion}: If one or more of the BJP\textsubscript{keywords} introduced in the previous section are present in the headline of a news article, we include the headline in B\textsuperscript{h}\textsubscript{yy-mm}. Similarly, if one or more of the Congress\textsubscript{keywords} introduced in the previous section are present in the headline of a news article then we include that headline in C\textsuperscript{h}\textsubscript{yy-mm}. If the headline contains keywords from both the sets of keywords, it is included in both the documents. 

\noindent\textit{Score of each document}: Each of the headlines forms one entity of attention to the general populace. So, we simply count the number of headlines included for each of the document as score of that document.

\noindent\textbf{Content} 

\noindent\textit{Sentence inclusion criterion}: If one or more of the BJP\textsubscript{keywords} are present in a sentence of the content of a news article, we include that sentence in B\textsuperscript{c}\textsubscript{yy-mm}. Similarly, if one or more of the Congress\textsubscript{keywords} are present in a sentence of the content of a news article, we include that sentence in C\textsuperscript{c}\textsubscript{yy-mm}. If the sentence contains keywords from both the sets of keywords, it is included in both the documents. 

\noindent\textit{Score of each document}: In contrast to the headlines, content is consumed by volume of words written. As words form the atomic unit of semantic expression, we use the number of words in the whole document as the score of that particular document in case of content metric.

\noindent\textbf{Point of view imbalance}

To understand which party's point of view is presented, we turn to the narrative verbs like ``say'' and ``tell''. Whereas, much research~\cite{ffgg} has gone into quote attribution, we find that the majority of point of views presented in the newspaper is indirect speech in reported form. So, to account for both, we count the number of times a noun phrase, containing at least one keyword/keyphrase of BJP or Congress, is the subject of a sentence. Thus, we will be able to get the sentences where something said by BJP or Congress or some member of the party has been highlighted.

\noindent\textit{Sentence inclusion criterion}: The sentence inclusion criterion is exactly similar to the one used for determination of content imbalance.

\noindent\textit{Score of each document}: We wish to get a rough statistical estimate of the number of words used to represent the point of view of each party by this score. So, as discussed above, we pick those sentences which contain any of the forms of the narrative verbs like ``say'' and ``tell'' as the main verb and also have a subject noun phrase containing the keywords/phrases related to the specific party of the document. We thus calculate the total number of words contained in the sentences picked from each document as the score of that document.

\subsection{Tonality imbalance} Here we discuss two different forms of imbalance metrics - (i) \textit{sentiment} imbalance and (ii) \textit{subjectivity} imbalance.

\noindent\textbf{Sentiment imbalance} 

We have done sentiment analysis for the articles using the widely used VADER sentiment analyzer of NLTK\footnote{\url{https://www.nltk.org/}}.

\noindent\textit{Sentence inclusion criterion}: The sentence inclusion criterion is exactly similar to the one used for coverage imbalance in content.

\noindent\textit{Score of each document}: Sentiment associations with keywords are studied here following Zhang et al~\cite{biascred}'s analysis of sentiment association with topics to determine imbalance/bias of a news-source. So, for each of the sentences in the BJP or Congress document, we determine the positive/negative sentiments expressed in that sentence using the VADER sentiment analyzer. Now, we get the average of the sentiments of these sentences weighted by the number of words of these sentences as the final score of the document. We use such weighting scheme to account for the semantic space as the sentiment is being expressed over the words for each sentence. So the density of the sentiments per word is a useful measure here.

\begin{figure*}[ht]
\centering 
\subfloat[Dendogram of coverage imbalance (headline)timelines clustering.]{\includegraphics[width=0.3\linewidth]{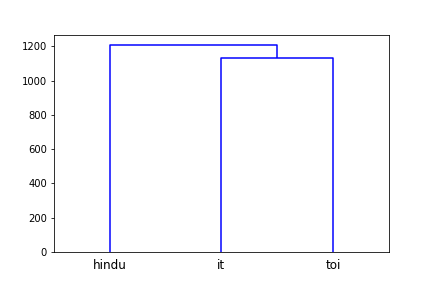}
\label{fig:1}} \hfil
\subfloat[Dendogram of coverage imbalance (content) timelines clustering.]{\includegraphics[width=0.3\linewidth]{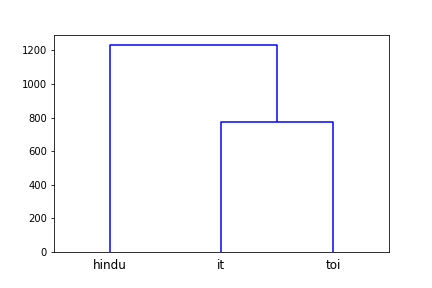}
\label{fig:2}}
\hfil
\subfloat[Dendogram of point of view imbalance timelines clustering.]{\includegraphics[width=0.3\linewidth]{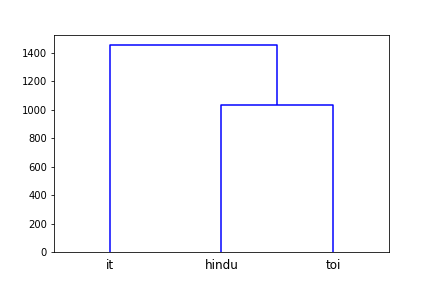}
\label{fig:3}}

\subfloat[Dendogram of positive sentiment imbalance timelines clustering.]{\includegraphics[width=0.3\linewidth]{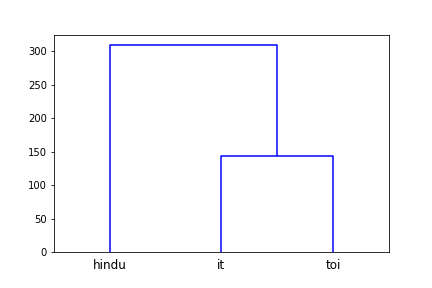}
\label{fig:4}}\hfil
\subfloat[Dendogram of negative sentiment imbalance timelines clustering.]{\includegraphics[width=0.3\linewidth]{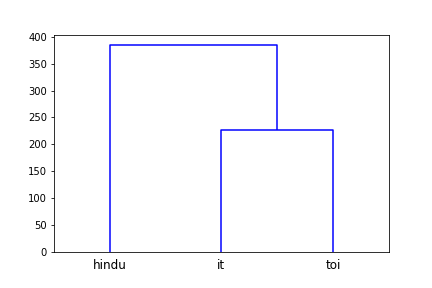}
\label{fig:5}}\hfil
\subfloat[Dendogram of subjectivity imbalance timelines clustering.]{\includegraphics[width=0.3\linewidth]{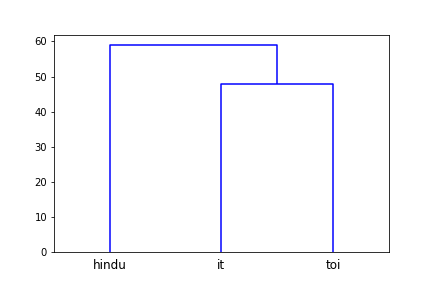}
\label{fig:6}}

\caption{Dendogram of different measures of imbalances across different newspapers}
\label{fig:dend1}
\end{figure*}

\begin{figure*}[h]
\centering
\subfloat[Dendogram of Times of India]{\includegraphics[width=0.3\linewidth]{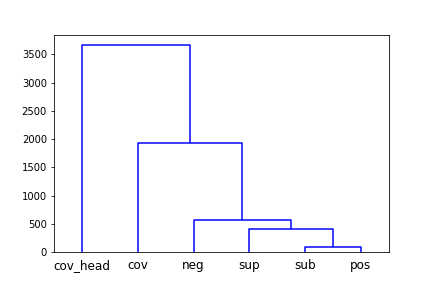}
\label{fig:s}} \hfil
\subfloat[Dendogram of India Today]{\includegraphics[width=0.3\linewidth]{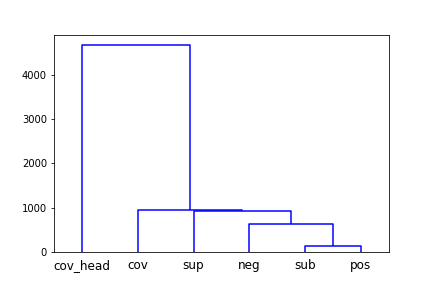}
\label{fig:s2}} \hfil
\subfloat[Dendogram of The Hindu]{\includegraphics[width=0.3\linewidth]{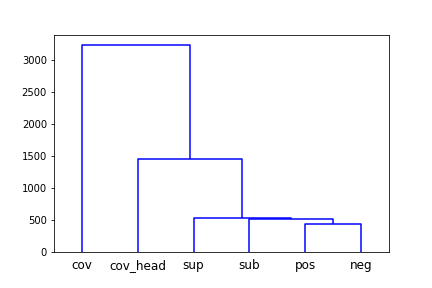}
\label{fig:s3}} \hfil
\caption{Dendogram of different newspapers across different measures of imbalances.(Acronyms used and their full forms:  cov\_h: coverage imbalance in headings; cov: coverage imbalance in content; sup: superlative and comparative imbalance; pos: positive sentiment imbalance; neg: negative sentiment imbalance; subj: subjectivity imbalance)}
\label{fig:dend2}

\end{figure*}

\noindent\textbf{Subjectivity imbalance}

We compute various quantities here most of which pertain to some notion of subjectivity as per the standard literature. In particular, we compute \textit{subjectivity} and uses of \textit{superlatives} \& \textit{comparatives} in the text of the content using the TextBlob\footnote{\url{https://github.com/sloria/textblob}} \& NLTK\footnote{\url{https://www.nltk.org/}} library. 

\noindent\textit{Sentence inclusion criterion}: For all the measures of subjectivity, we use the same sentence inclusion criterion as the one used for coverage imbalance in content.

\noindent\textit{Score of each document}: Sentiment and subjectivity are very related concepts. So, we use the same rationale of scoring here as used in the section of sentiment imbalance. For all these subjectivity metrics, for each of the sentences in the document we determine the score of the sentences using the aforementioned tools. Now, we obtain the average of these scores across all the sentences weighted by the number of words of these sentences as the final score of the document for each metric using the same rationale as described in the previous section.

For superlatives and comparatives, we count the average percentage of superlatives and comparatives used in the sentences where any of the party is mentioned (based on the same \textit{keyword} based filtering), as an alternative measure of subjectivity. 

\subsection{Summary based on time series clustering}

For each individual imbalance metric and each newspaper, one can obtain a time series of the directed scores spanning over 9+ years. While one can always look into each such time series data to make an inference, our idea was to look for universal characteristics of imbalance across the three news outlets. To this end, we cluster the time series of imbalance scores using the standard dynamic time warping (DTW) approach. We use hierarchical agglomerative clustering to understand the similarity among the newspapers based on their temporal imbalance characteristics. In addition, this clustering technique also summarizes which of the metrics have remained closer to each other over time.

\subsection{Summary based on aggregation of scores}

We aggregate the documents of each of the two parties over the whole timeline to obtain two aggregate documents. Next we compute each of the above metrics using the equation~\ref{e1}, to obtain the aggregate imbalance score corresponding to each metric and each political party.

\subsection{Word embedding association test (WEAT)}
In order to understand how the popularity of the political parties among the people of India has changed over time we calculate the year specific word embeddings on our corpus using methods used to measure semantic shift in words over time. 

Previous research\cite{xpol} suggests that frequently used words have the least shift of their meaning over time. Hence we use top 1000 words in our corpus (excluding the words \textit{BJP} and \textit{Congress} consciously) to align the word embeddings trained for different time periods. We train word2vec~\cite{Mikolov2013EfficientEO} with SGNS (skip-gram with negative sampling), to create embeddings for each of the year in our dataset. Let $W(t) \in \mathbb{R}^{d \times V}$ be the matrix of word embeddings learnt for year $t$ and for vocabulary $V$. Following Hamilton et al~\cite{hamilton}, we jointly align the word embeddings while generation, using the top 10000 common words present across time periods $t_1$ and $t_2$ by optimizing:

\begin{equation*}
 R^t=\argmin_{Q_{t_1t_2}^TQ_{t_1t_2}=I} ||QW^{t_1} -W^{t_2}||
\end{equation*}

For simplicity we assume $Q_{t_1t_2}=I \forall t_1, t_2$. After alignment, we measure the WEAT score of the words \textit{BJP} and \textit{Congress} with the opposite set of words $A_1=$\{good, honest, efficient, superior\} and $A_2=$\{bad, dishonest, inefficient, inferior\} using the algorithm presented in~\cite{brunet19a}. 

The differential association of a word $c$ with word sets $A_1$ and $A_2$ is given by

\begin{equation*}
  g(c, A_1, A_2, w) = \tiny{\begin{array}{c}mean \\ a \in A_1\end{array}}  \cos{(w_c, w_a)} - \tiny{\begin{array}{c}mean \\ b \in A_2\end{array}} \cos{(w_c, w_a)}
\end{equation*}

where $w$ is the set of word embeddings, $w_x$ is the word embedding for the word $x$.

Now, the WEAT score is calculated as

\begin{equation*}
  B_{weat}(w)=\frac{g'(s_1) - g'(s_2) }{\tiny{\begin{array}{c}SD \\ s_3 \in S_1,S_2\end{array}}  g(s_3 , A_1, A_2, w)} 
\end{equation*}

where
\begin{equation*}
  g'(s_i) = \tiny{\begin{array}{c}mean \\ s_i \in S_i\end{array}}  g(s_1 , A_1, A_2, w)
\end{equation*}

Here the word sets $S_1$ and $S_2$ are the keywords related to the political parties, as already discussed previously.



\section{Experiments and results}

In this section we discuss the key experiments and then detail the corresponding results.

\subsection{Summary based on time series clustering}

\begin{figure*}[h]
\subfloat[Content coverage trends.]{\includegraphics[width=0.45\linewidth]{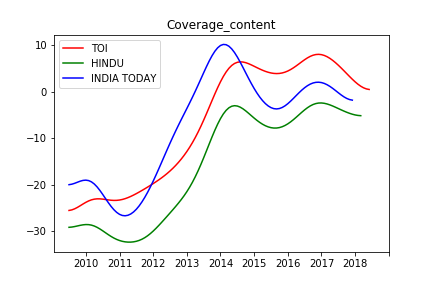}
\label{fig:covh}} 
\hfil
\subfloat[Positive sentiment trends.]{\includegraphics[width=0.45\linewidth]{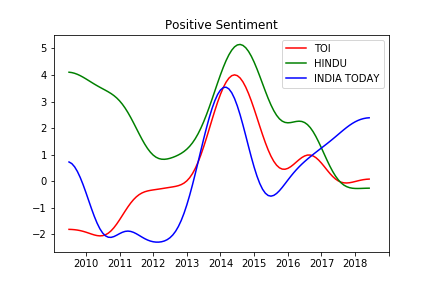}
\label{fig:p}} 
\caption{\label{fig:trend_temp}Temporal variation of imbalance in coverage of content and positive sentiments in the news articles for the different media houses.}
\end{figure*}

We have seven different imbalance metrics namely -- headlines coverage, content coverage, point of view, positive sentiment, negative sentiment, subjectivity and superlatives/comparatives. For a given news outlet therefore we shall have seven corresponding time series each spanning over 9+ years. Since there are three major news outlets in our dataset we have 21 time series in all. We cluster these 21 time series using DTW as discussed in the previous section. As evident from the results presented in the form of dendograms in Figure~\ref{fig:dend1}, \textit{Times of India and India Today exhibit stronger clustering across almost all the imbalance measures pointing to an interesting universal characteristic}. In order to delve deeper into the dynamics, we present in Figures~\ref{fig:trend_temp}(a) and~\ref{fig:trend_temp}(b) respectively, two representative time series plots of imbalance scores -- content coverage imbalance and positive sentiment imbalance.

\if{0}
\begin{figure}[h]
\centering 
\subfloat[Positive sentiment trends.]{\includegraphics[width=0.45\linewidth]{vader_pos}
\label{fig:vp}} \hfil

\subfloat[Negative sentiment trends.]{\includegraphics[width=0.45\linewidth]{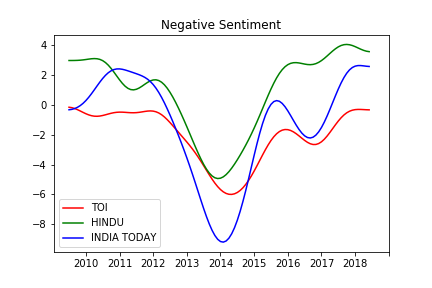}
\label{fig:vn}} 

\caption{\label{fig:trend_p}Temporal variation of coverage and positive sentiment imbalance in the news articles for the different media houses.}
\end{figure}
\fi
\noindent\textit{Content coverage imbalance}: In Figure~\ref{fig:trend_temp}(a), we plot the directed imbalance scores of content for each of the media houses over time. The first noticeable trend is that the media houses have distinct relative bias very consistent over the timeline. The Hindu has been especially consistent in maintaining a 5-10\% shift in coverage toward the Congress party than the other two news organizations. Consequent to relatively higher leaning of The Hindu in coverage of the Congress party, The Hindu always remained Congress leaning (below zero in the curve) unlike its two peers. The shift between TOI and India Today is not that apparent thus providing the empirical justification of the results obtained from the DTW clustering.

\noindent\textit{Positive sentiment imbalance}: In Figure~\ref{fig:trend_temp}(b), we observe that The Hindu has an imbalance score higher than the zero mark throughout till 2017 showing higher density of positive sentiments toward BJP. This behavior is in drastic contrast with the other two news outlets thus providing the justification in support of the DTW results. Of course, the election year (2014, which also marked significant change in vote share and public sentiment) observes high positive sentiments in favor of BJP in general.

As a next step, we cluster the seven different time series corresponding to the respective imbalance metrics for each of the news outlets separately (see Figure~\ref{fig:dend2}). For all the news outlets we again observe \textit{a universal pattern whereby the tonality based imbalances are clustered more strongly exhibiting their distinctions with the coverage and point of view imbalances}. 


\subsection{Summary based on aggregation of scores}

Table~\ref{_agg} shows the aggregate imbalance scores for each of the newspapers across all metric for a fair quantitative comparison. We can see that there is no clear winner or loser in terms of imbalances. TOI reports the highest degree of imbalance in case of three metrics and The Hindu \& India Today show highest imbalance in case of two metrics each. One peculiar point to note here is that for The Hindu, the average positive and negative sentiments are both BJP leaning. One can argue that this is counterintuitive since the density of positive sentiments in favor of one party being high for a media house in one month should imply that the density of negative sentiments has to be low in favor of that party for that month. Although intuitive, this is not obvious as both positive and negative sentiments can be expressed highly about a party if more about that party is discussed in the inkspace. 
For instance, at multiple time points, both the positive and negative sentiment scores for The Hindu are much above the zero line unlike the other two media houses (data not shown for paucity of space).



\begin{table}[t]
\centering

\begin{tabular}{llll}
\hline
 & TOI  & The Hindu & India Today \\ \hline
 CovHead & $\uparrow$\textbf{16.18} & $\uparrow$9.87 & $\uparrow$14.72  \\ \hline 
 CovCon & $\downarrow$4.49 & $\downarrow$\textbf{11.72} & $\downarrow$2.50  \\ \hline 
 PoV & $\uparrow$\textbf{78.31} & $\uparrow$70.14 & $\uparrow$62.45  \\ \hline 
 PosSent & $\uparrow$0.44 & $\uparrow$\textbf{2.42} & $\uparrow$0.20  \\ \hline 
 NegSent & $\downarrow$\textbf{2.02} & $\uparrow$1.05 & $\downarrow$0.54  \\ \hline 
 Subj & $\downarrow$0.28 & $\downarrow$0.57 & $\uparrow$\textbf{1.04}  \\ \hline 
 SupComp & $\downarrow$1.15 & $\downarrow$1.96 & $\downarrow$\textbf{5.22}  \\ \hline

\end{tabular}%

\caption{\label{_agg} Aggregate absolute imbalance scores: An upward arrow at the left of any number denotes an imbalance toward BJP and vice versa. The highest absolute imbalance score for each metric has been highlighted in boldface.} 
\end{table}

\begin{figure}[t]
\centering

\includegraphics[width=0.9\linewidth]{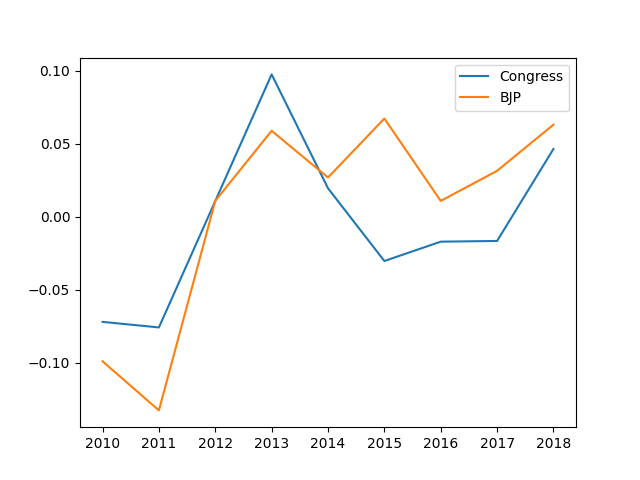}
\caption{Temporal variation of \textit{popularity} of each party for the India Today corpus. The trends are very similar for the other two news outlets.}\label{fig:weat1}
\hfil
\end{figure}

\begin{figure*}[ht]
\centering
\subfloat[Inverse of standard deviation (imbalance in coverage of all states together)]{\includegraphics[width=0.32\linewidth]{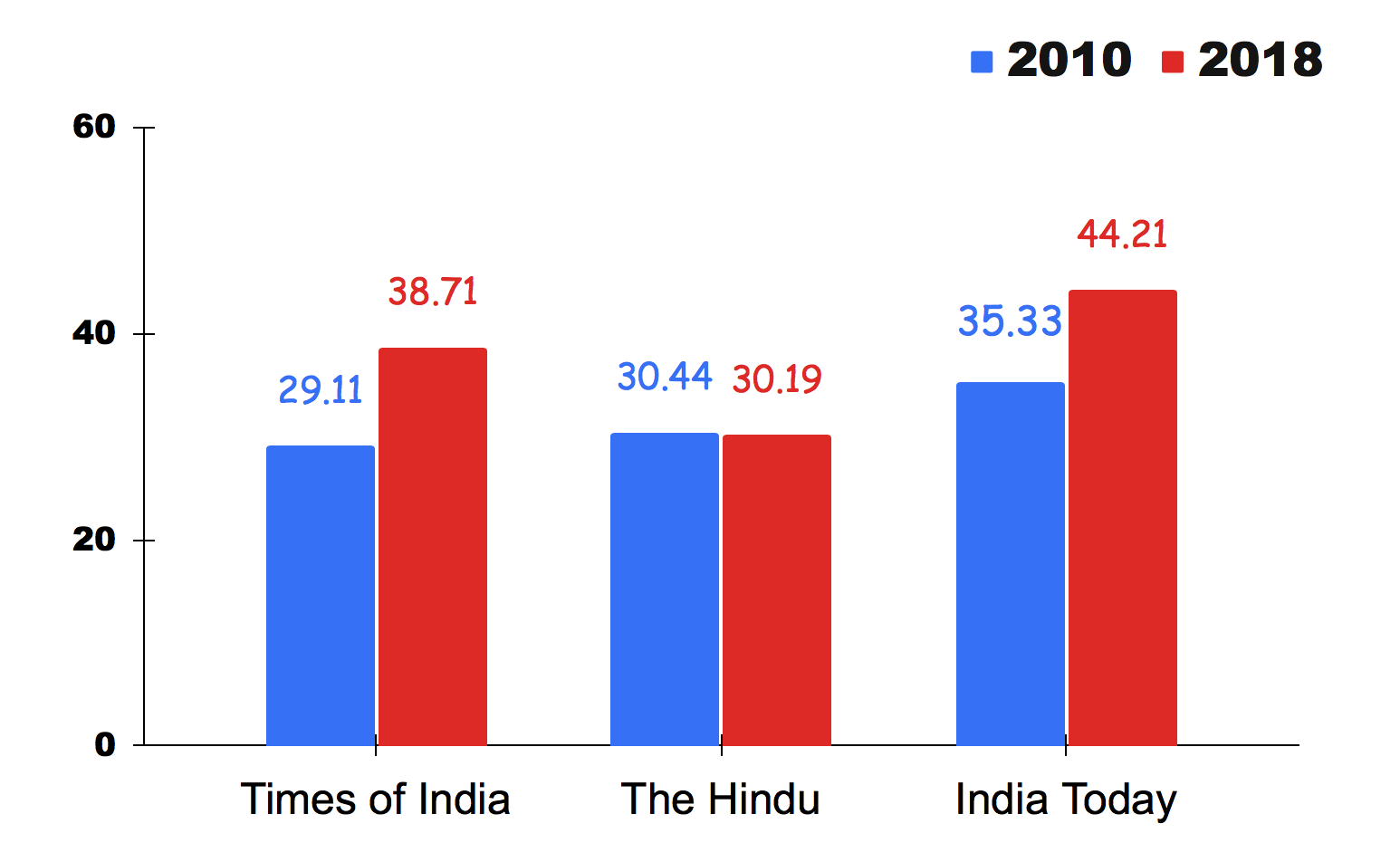}
\label{fig:sub1}}\hfil
\subfloat[Coverage (in \%) of bottom 20\% states]{\includegraphics[width=0.32\linewidth]{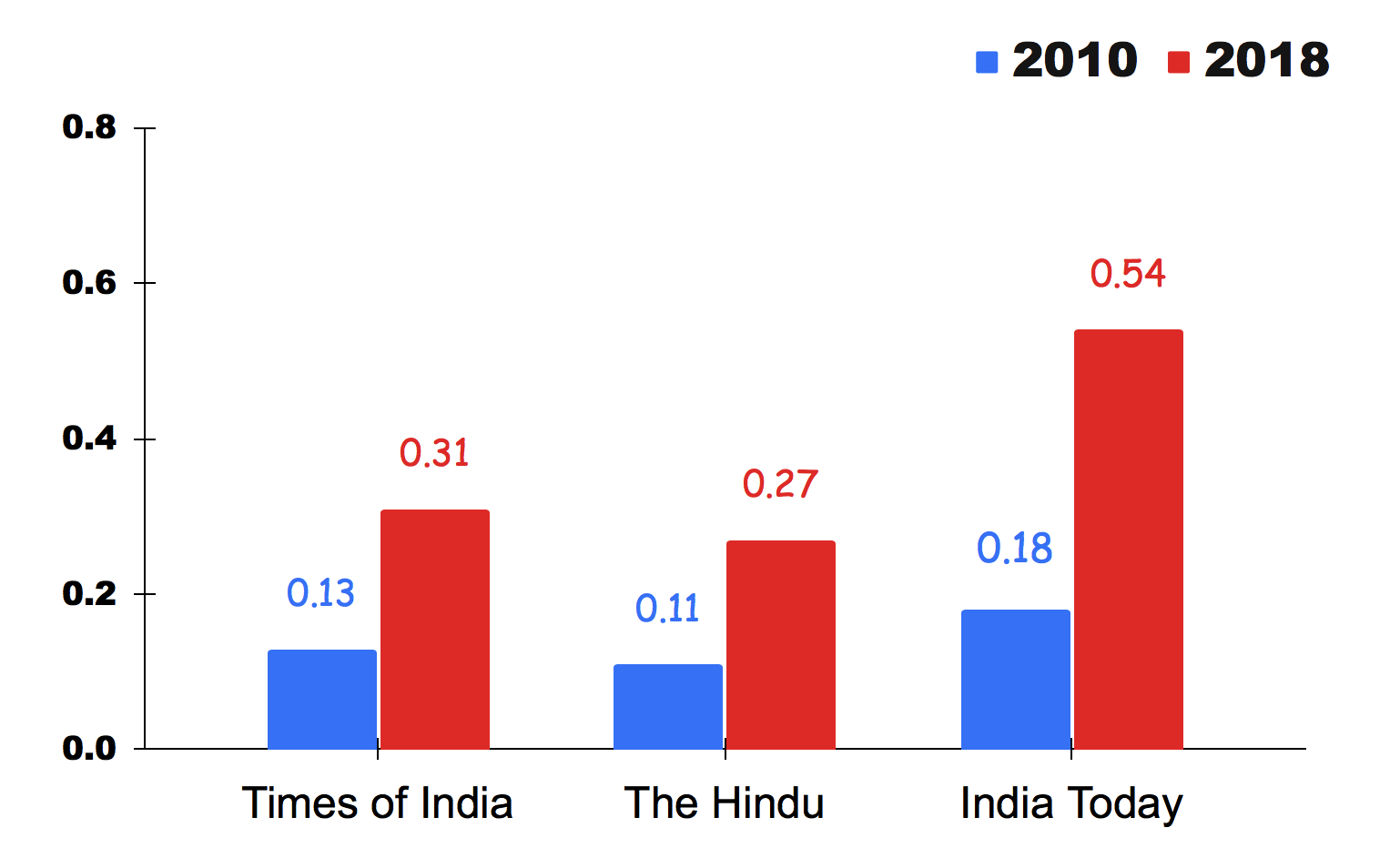}
\label{fig:sub2}}\hfil
\subfloat[Coverage (in \%) of bottom 50\% states]{\includegraphics[width=0.32\linewidth]{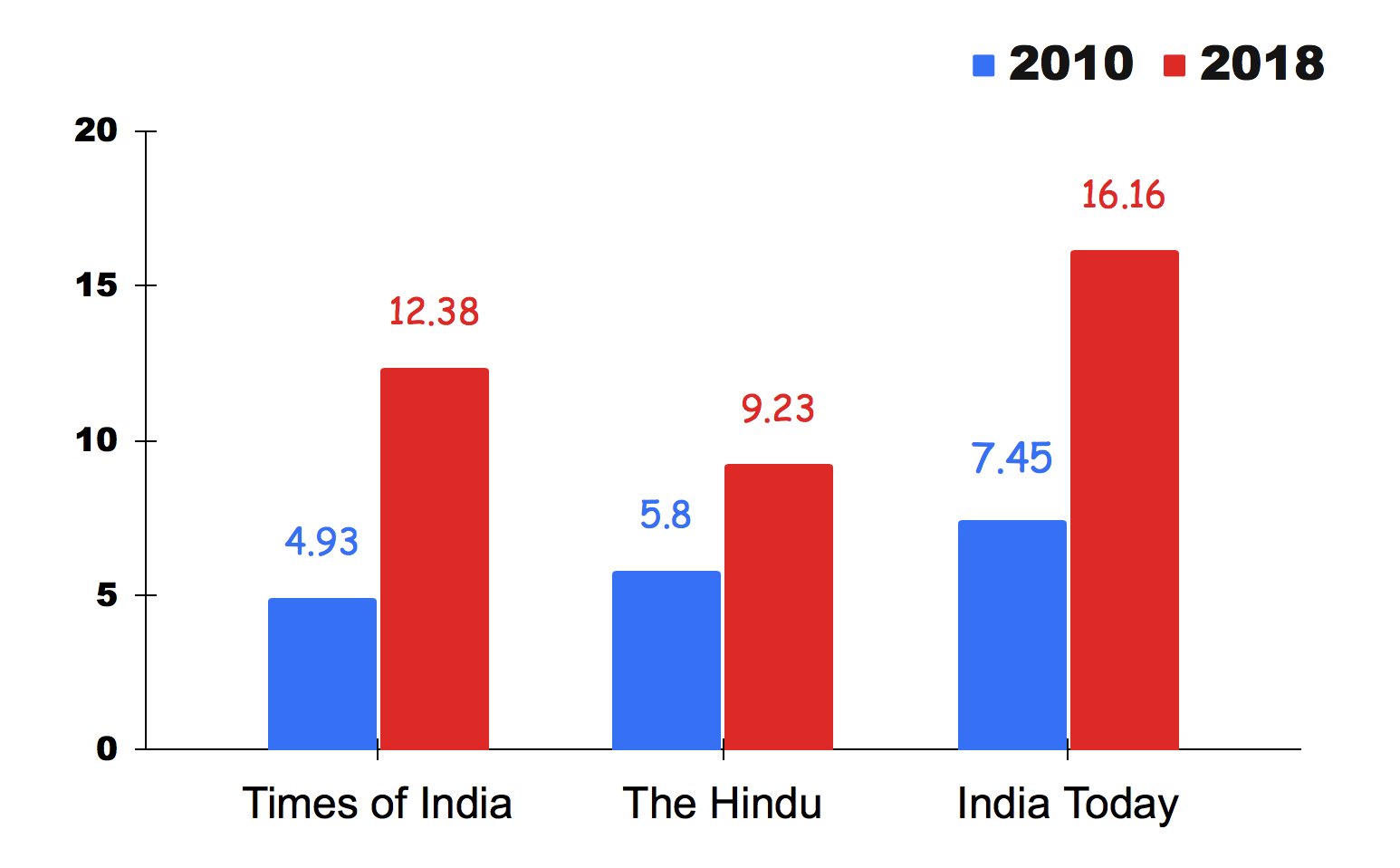}
\label{fig:sub3}}

\caption{\label{std_fig}Trends in imbalance in coverage of states over time.}
\end{figure*}

\subsection{WEAT scores to determine party popularity}\label{pop_sec}
We calculate the differential association over time and plot that over the years. We use this distance as a proxy for popularity as portrayed by that particular news media. From Figure~\ref{fig:weat1}, it is evident that \textit{BJP} gained popularity in news very fast post 2011, surpassing popularity of \textit{Congress} in 2014, the year of legislative assembly election when incumbent \textit{BJP} overthrew the ruling Congress government. We also see the popularity of \textit{Congress} increasing again since 2016, the year of demonetization, that possibly had a strong impact on the economy of India and specially on the poorest ones of the country.\footnote{\url{https://www.bbc.com/news/world-asia-india-46400677}}



\if{0}
\begin{figure*}[ht]
\centering
\begin{subfigure}{0.64\columnwidth}
  \centering
  \includegraphics[width=0.95\linewidth]{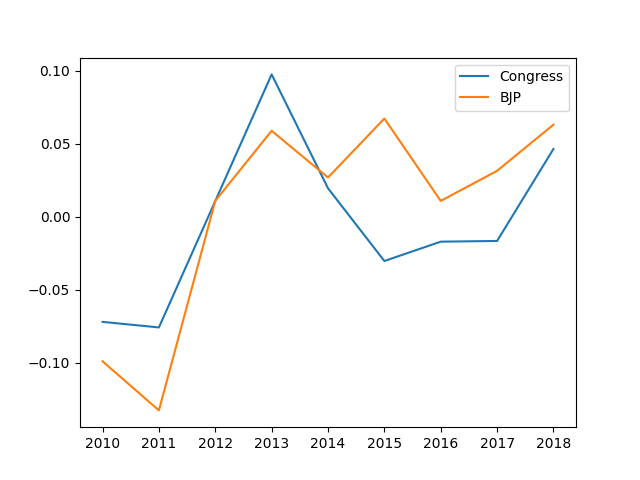}
  \caption{Times of India}
  \label{fig:sub1}
\end{subfigure}
\begin{subfigure}{0.64\columnwidth}
  \centering
  \includegraphics[width=0.95\linewidth]{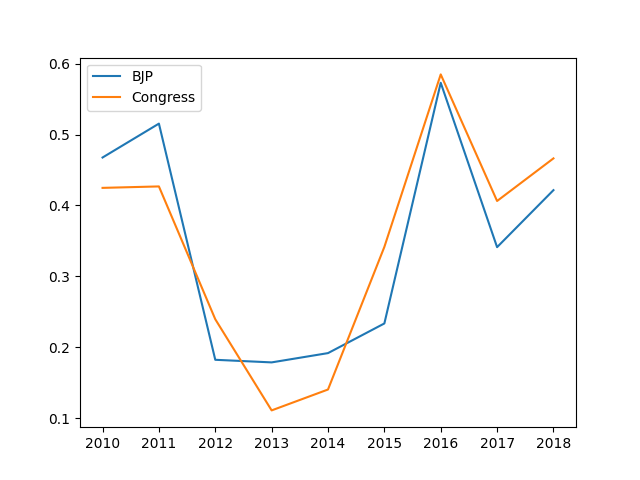}
  \caption{India Today}
  \label{fig:sub2}
\end{subfigure}
\begin{subfigure}{0.64\columnwidth}
  \centering
  \includegraphics[width=0.95\linewidth]{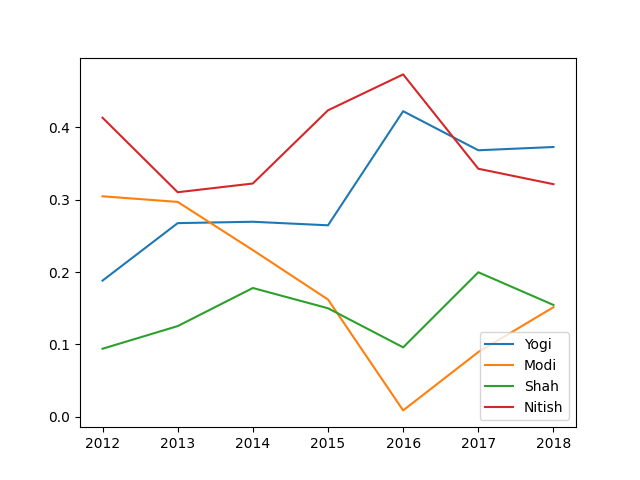}
  \caption{The Hindu}
  \label{fig:sub3}
\end{subfigure}%
\caption{\label{weat_fig}Temporal variation of \textit{popularity} of each party as measured from WEAT score}
\end{figure*}
\fi

\subsection{Imbalance in portrayal of different states/cities}

Non-Hindi speaking states and especially states from north-east India\cite{giri2015content}, Jammu \& Kashmir have often alleged other parts of India of cultural exclusion\footnote{\url{https://towardfreedom.org/story/indian-medias-missing-margins/}}.
We attempt to understand how much of those allegations are true and if the situation is changing over time.
\subsubsection{City level analysis}
\begin{figure}[h]
\centering
\includegraphics[width=0.9\columnwidth]{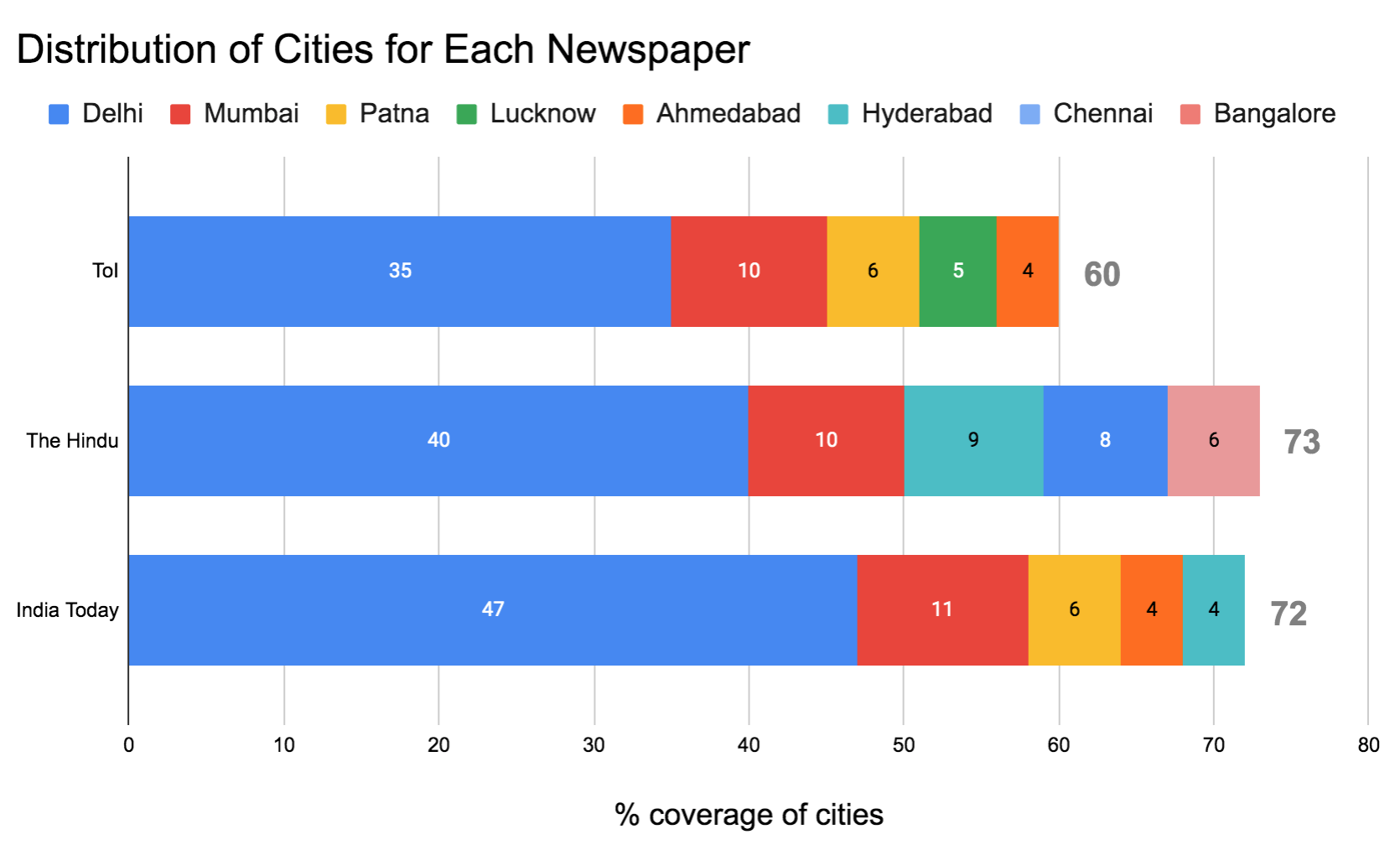}
\caption{\label{figlocimbalance} Coverage of different cities by the three news outlets (only top five cities covered are shown to facilitate increased visibility).}
\end{figure}

We prepare a list of 25 most populous cities in India according to the census report of 2011\footnote{\url{http://censusindia.gov.in/2011-Common/CensusData2011.html}\label{foot6}} and measure how these cities are covered by the news articles. We take one entry of a city if the city is mentioned at least once in a news article. We thus calculate the share of each city for a specific media house. We illustrate this imbalance in the coverage of cities in Figure~\ref{figlocimbalance}. 

\begin{figure*}[htb]
\centering
\subfloat[Times of India, all 9 years]{\includegraphics[width=0.32\linewidth]{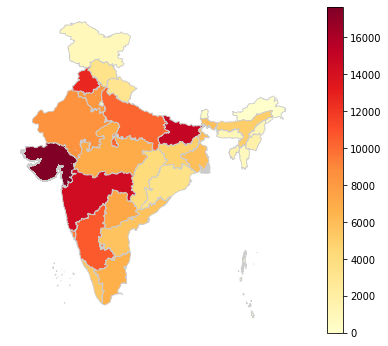}
\label{fig:s}}\hfil
\subfloat[India Today, all 9 years]{\includegraphics[width=0.32\linewidth]{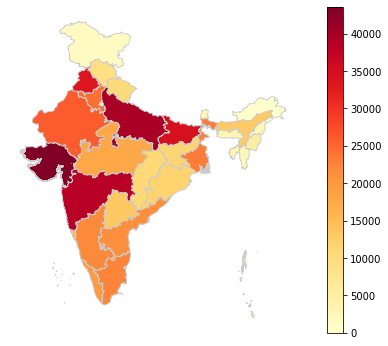}
\label{fig:sc}}\hfil
\subfloat[The Hindu, all 9 years]{\includegraphics[width=0.32\linewidth]{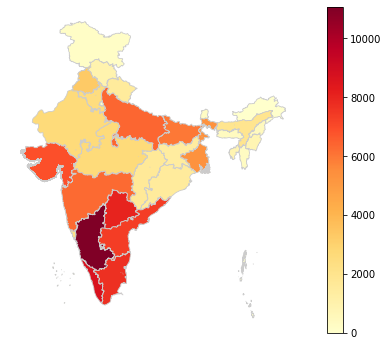}
\label{fig:sc}}

\subfloat[Times of India, 2018]{\includegraphics[width=0.32\linewidth]{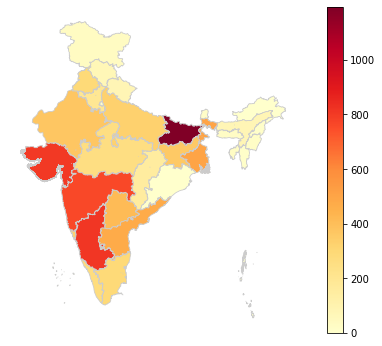}
\label{fig:sc}}\hfil
\subfloat[India Today, 2018]{\includegraphics[width=0.32\linewidth]{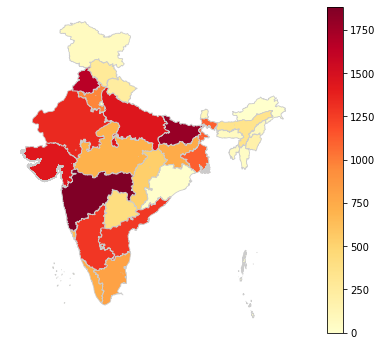}
\label{fig:sc}}\hfil
\subfloat[The Hindu, 2018]{\includegraphics[width=0.32\linewidth]{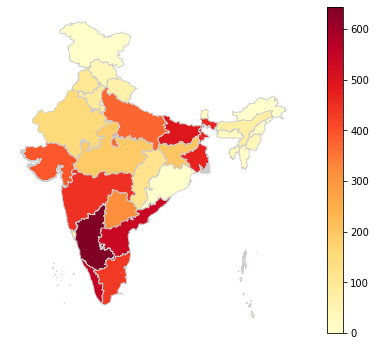}
\label{fig:sc}}

\subfloat[Times of India, 2018]{\includegraphics[width=0.32\linewidth]{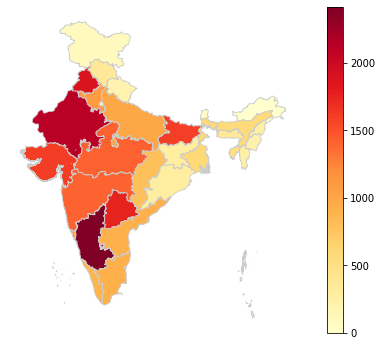}
\label{fig:sc}}\hfil
\subfloat[India Today, 2018]{\includegraphics[width=0.32\linewidth]{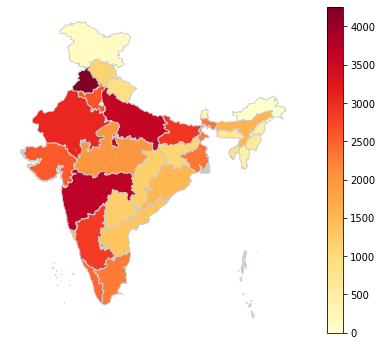}
\label{fig:sc}}\hfil
\subfloat[The Hindu, 2018]{\includegraphics[width=0.32\linewidth]{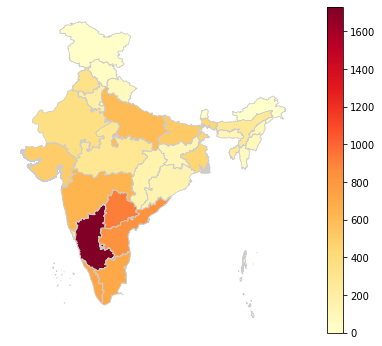}
\label{fig:sc}}

\caption{Number of articles mentioning each state for different newspapers across different time periods}
\label{fig:loc2}

\end{figure*}

\noindent\textit{Observation}: We can see from Figure~\ref{figlocimbalance} that the distribution of cities covered by each of the newspapers is heavily skewed with the most frequently covered five cities corresponding to 60-70\% of the articles. Delhi and Mumbai are the two most dominant cities on this list for all the three newspapers.

\subsubsection{State level analysis}
Now we attempt to understand if the situation is similar in state level and if \textit{yes}, then which states are covered poorly.
We collect all the state names from the census report of 2011\textsuperscript{\ref{foot6}} and search for their occurrence across the corpus. We note the number of articles a specific state is mentioned in and plot the same. We do this experiment for all the three newspapers and for each newspapers, we once plot for only 2010, once only for 2018 to understand the evolving trend.

\noindent\textit{Observations}: It is evident from Figure~\ref{fig:loc2} that the allegations mentioned in the start of the section is true. The states of Jammu \& Kashmir, states in the north-east and some non-Hindi speaking states like Orissa or Jharkhand are squarely ignored by all the three newspapers. Hindu seems to stand out in coverage of states from the other two newspapers covering mostly south Indian states. Looking at the maps comparatively from 2010 to 2018, it seems that the situation is improving and more states are getting covered by the national newspapers over time though equality among the states may be a long way.

\subsubsection{Is the situation getting better/worse?}
To better understand the trend, we attempt to quantify homogeneity using three metrics. First we convert the counts of states to probabilities by dividing them with the total number of mentions of all cities. Next we define the first metric of homogeneity as the inverse of the standard deviation of the probability distribution (since standard deviation is an established measure of homogeneity). The next two measures take a more boxed approach trying to understand if the low priority states for that particular newspaper is getting higher attention over time. Here low priority states for the second metric is the bottom 20\% states and for the third metric is the bottom 50\% states.

\noindent\textit{Observations}: From Figure~\ref{std_fig}, we observe that the intuitive conclusions drawn from the last analysis stands true for \textit{Times of India} and \textit{India Today} for all the metrics. For the Hindu in the first metric (i.e., inverse of standard deviation) we see no clear trend. However, in both the other metrics we note that the newspaper is clearly diversifying its coverage over the states.

\section{Further insights}

In order to obtain further insights, we perform a \textit{cloze} task~\cite{10}, i.e., a task that requires completion of a sentence by correctly predicting the masked/hidden word. For instance, in the following cloze task -- ``Sun is a huge ball of $\langle  mask \rangle$, ``fire'' is a likely completion for the missing word. Given a cloze test, well-known language models like RoBERTa~\cite{liu2019roberta}, produce a sequence of tokens with their corresponding probabilities to fill the given blank in the input sentence. We train RoBERTa (initialized with RoBERTa-base~\cite{liu2019roberta}) for each of our newspapers for each year present in the corpus separately for 20000 iterations following the language model training procedure described in Khalidkar et al~\cite{khadilkar2021gender}. This results in a total $3*9=27$ different models.
We use these models (representative/mouthpiece of each newspaper at different times of the 9 years in our corpus) to answer the following three questions --
(a) can one track the changing priorities for India as depicted by each news media house? (b) how are these newspapers reporting popularity of one party over the other, for these 9 years? and (c) how are newspapers presenting perception about the Indian economy?

\if{0}\noindent{\textbf{Analysis:}} From the results of Cloze task 1, depicted in Table ~\ref{tab:cloze_table}, we can see that 2018 shows an increase in global locations with names of countries mentioned around the globe whereas it sees a decrease in names of specific cities and Indian locations such as Indian states and cities. This pattern is very consistent across newspapers. Also, most of the foreign locations mentioned to be decreasing in 2018 like Pakistan, Washington, England and Nepal are close allies of India. We showed these results to 9 Indians in verse with the events in Indian government. They all unanimously pointed out that the Indian Prime minister of the term 2014-2019 made many more visits to new countries in order to expand India's presence and influence in global diplomacy than his precedent, the prime minister of the term 2009-2014 who focused mostly on Indian affairs and neighboring countries and allies\footnote{\url{https://www.altnews.in/narendra-modi-vs-manmohan-singh-on-foreign-visits-fact-checking-amit-shahs-claim/}}. In conclusion they all agreed that the results of the Cloze task are very reasonable.
One unintuitive pick by RoBERTa for top decreasing token in 2018 for this cloze task was attendance and indeed because of the frequent foreign trips to different countries, attendance of the prime minister in the Legislative Assembly was low in that term than the previous term.\footnote{\url{https://www.indiatoday.in/india/story/nation-wants-to-see-more-of-pm-modi-in-parliament-1288793-2018-07-18}}\fi

\begin{table*}[ht]
\scriptsize
\centering
\begin{tabular}{|p{25mm}|p{20mm}|p{20mm}|p{20mm}|p{20mm}|p{20mm}|p{20mm}|}
\hline
Probe &
  \multicolumn{2}{l|}{RoBERTa$_\textrm{(TOI)}$} &
  \multicolumn{2}{l|}{RoBERTa$_\textrm{(Hindu)}$} &
  \multicolumn{2}{l|}{RoBERTa$_\textrm{(IT)}$} \\ \hline
 &
  \multicolumn{2}{l|}{2010 $\longrightarrow$ 2018} &
  \multicolumn{2}{l|}{2010 $\longrightarrow$ 2018} &
  \multicolumn{2}{l|}{2010 $\longrightarrow$ 2018} \\ \hline
 &
  $\uparrow$ 2018 &
  $\downarrow$ in 2018 &
  $\uparrow$ 2018 &
  $\downarrow$ 2018 &
  $\uparrow$ 2018 &
  $\downarrow$ 2018 \\ \hline
\if{0}$Cloze_1$ &
  Dubai, France, Singapore, China, Italy, Switzerland, Pakistan, Germany, Malaysia, Paris, Hawaii, Japan, Delhi &
  Chennai, India, Washington, Mumbai, Nepal, Beijing, Australia, Bangalore, England, Islamabad, town, Bihar, attendance, London &
  Singapore, London, Brussels, jail, China, Dubai, France, Bahrain, England, Berlin, Australia, hospital, Germany, Tokyo &
  Islamabad, Delhi, Washington, India, Beijing, Riyadh, Rome, Pakistan, Kabul, Mumbai, Malaysia, Chennai, Baghdad, Nepal &
  Singapore, Italy, Finland, Germany, India, Russia, Switzerland, Malaysia, France, Japan, Jakarta, Rome, China, England, Paris, Brussels &
  London, Islamabad, Delhi, Washington, Beijing, Pakistan, Australia, Toronto, Chennai, Bangkok, Mumbai, Dubai, Canada \\ \hline  \fi
  
The main issue in India is $\langle$ \textit{mask} $\rangle$ &
  unemployment, water, terrorism, jobs, farmers, corruption, employment, women, agriculture, reservation, fuel, caste, GST, development, food &
  Kashmir, migration, terror, Afghanistan, security, India, democracy, Pakistan, elections, insecurity, peace, violence, inflation &
  unemployment, corruption, employment, water, GST, development, jobs, reservation, agriculture, poverty, immigration, housing, governance, money, food &
  Kashmir, terrorism, terror, prices, India, Pakistan, trade, Afghanistan, inflation, peace &
  unemployment, corruption, poverty, education, GST, water, Aadhaar, malnutrition, pollution, agriculture, farmers, immigration, inequality, healthcare, democracy &
  terrorism, Kashmir, terror, Afghanistan, security, Pakistan, trade, inflation, development \\ \hline
The economy of India is $\langle$ \textit{mask} $\rangle$ &
  growing, strong, slowing, weak, stagnant, thriving, developing, intact, poor, healthy, deteriorating, flourishing, bleeding, backward, expanding &
  crumbling, shrinking, dying, suffering, divided, rotting, exploding, changing, broken, paralyzed, declining, collapsing, reeling, weakening, fragile &
  shrinking, crumbling, dead, collapsing, destroyed, slowing, falling, broken, bleeding, different, intact, deteriorating, poor &
  growing, vibrant, stagnant, flourishing, struggling, strong, booming, recovering, healthy, thriving, weak, improving, stable, fragile, sound &
  changing, suffering, shrinking, developing, huge, dying, broken, transforming, declining, great, poor, weak, stagnant, flourishing &
  growing, robust, booming, slowing, improving, fragile, contracting, thriving, strong, evolving, stable, good, expanding, recovering, vibrant \\ \hline
\end{tabular}
\caption{Top tokens increasingly and decreasingly accepted as answer in 2018 for the cloze task (a) \& (c).}
\label{tab:cloze_table}
\end{table*}

\begin{figure*}[t]
\centering
\subfloat[Times of India]{\includegraphics[width=0.32\linewidth]{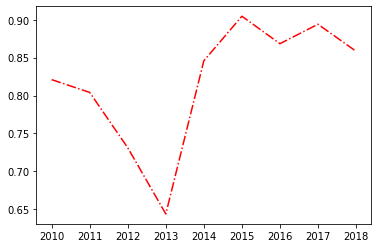}
\label{fig:pop1}} \hfil
\subfloat[India Today]{\includegraphics[width=0.32\linewidth]{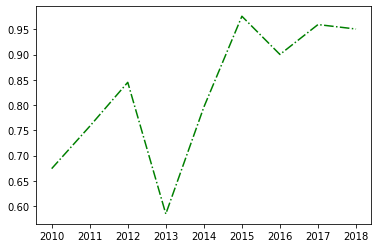}
\label{fig:pop2}} \hfil
\subfloat[The Hindu]{\includegraphics[width=0.32\linewidth]{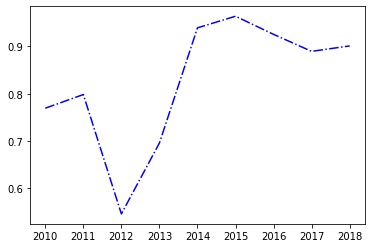}
\label{fig:pop3}}

\caption{\label{popularity_fig}Popularity of BJP over Congress quantified from the results of Cloze test 4, plotted over the years}
\end{figure*}

\subsection{Can one track changing priorities?} To understand the changing priorities of India as a nation over the last decade, we propose the following cloze task query -- ``The main issue in India is $\langle   mask \rangle$''. We attempt to understand how RoBERTa's answer changes for this specific query from 2010 to 2018.
To this purpose, we take a union of top 50 tokens given as output for RoBERTa$_{2010}$ and RoBERTa$_{2018}$. We then rank the top tokens which underwent maximum positive change from 2010 to 2018 as an answer to the cloze test (i.e. the tokens which are more accepted as answer in 2018 than in 2010 for the cloze test). We also rank the top tokens which underwent maximum negative change from 2010 to 2018 as an answer to the cloze test (i.e. the tokens which are less accepted as answer in 2018 than in 2010 for the cloze test). We show maximum of 15 such tokens in order of probability (higher to lower).

\noindent\textit{Analysis and observations}: From Table~\ref{tab:cloze_table}, we see a similar pattern reverberating across the news media houses.
The focus of India in 2018 is more on economic issues like \textit{unemployment}, \textit{jobs}, \textit{corruption}, \textit{poverty}, \textit{GST}, \textit{food} and \textit{reservation} and less on border issues like \textit{Kashmir}, \textit{Pakistan}, \textit{Afganistan} and \textit{security}. More basic demands like \textit{food}, \textit{housing}, \textit{water} and \textit{agriculture} are popping up in 2018.  We showed these results to 9 Indians in verse with the events in Indian government. All of them unanimously agreed that these are due to the changing landscape of events affecting India from 2010 to 2018. The period 2008 -- 2010 saw a lot of coordinated bombing and shooting attacks by terrorists on Mumbai, the economic capital of India resulting in mass killings and injuries. These issues mainly related to the India-Pakistan border conflicts emerge in the words popping up in the 2010 newspapers. Between 2014 -- 2018, on the other hand, India saw various economic reforms in the form of introduction of GST, demonetization, stress on online transactions and implementation and linking of AADHAR (an unified database of citizens like social security number in US) with banking for continuation of banking services. All these together led to the increase of priority of economy related words in these news outlets.


\subsection{How are these newspapers reporting popularity?}
We attempt to understand how popularity of one party over the other is reported in these newspapers and how they are similar or different from each other.
We define voting preference toward a specific political party $\langle   p \rangle$, $\forall \langle   p \rangle  \in $\{\textit{``BJP'',``Congress''}\}.  as:

\begin{equation}
V_{pop}(\langle   p \rangle ) = P_{RoBERTa}(\langle   mask \rangle  = \langle   p \rangle  | input =prompt)
\end{equation}
where
\begin{equation}
prompt=\text{``This election people will vote for } \langle   mask \rangle \text{.''}
\end{equation}

Further, we normalize these values to probabilities toward any of the two parties, arbitrarily selected to be \textit{BJP} (plotting both is redundant as $p_{congress}=1-p_{bjp}$) as

\begin{equation}
Pr_{pop}(``BJP'')=\frac{V_{pop}(``BJP'')}{V_{pop}(``BJP'')+V_{pop}(``Congress'')}
\end{equation}

\subsection{How are these newspapers reporting economic prosperity?}
We attempt to understand how media houses are reporting economic prosperity of India over time. Using the probe ``The economy of India is $\langle$ \textit{mask} $\rangle$'', we report the most probable outputs in ~Table~\ref{tab:cloze_table}

\vspace{2mm}

\noindent\textbf{\textit{Observations}}: We plot the probabilities in favor of ``BJP'' over time for each news media house in Figure~\ref{popularity_fig}. We observe that once again all the news media groups show a very similar pattern with the period 2012-2013, a year before the national election, to be the inflection point. The opposition `BJP' could defeat the incumbent `Congress' government with a large margin following gain in popularity in 2010-2011 largely due to corruption charges against the `Congress' which resulted in nationwide protest and a very influential anti-corruption movement in the capital. We see the popularity of `BJP' with respect to `Congress' only rose in the years following the election which seems intuitive as `BJP' won the 2019 election also with huge majority and increased vote share than 2014\footnote{\url{https://en.wikipedia.org/wiki/Results_of_the_2019_Indian_general_election}}.
Also, an interesting observation is that a huge policy failure like demonetization which arguably influenced the fall of GDP due to extreme shrinkage of money in circulation\footnote{\url{https://www.theguardian.com/world/2018/aug/30/india-demonetisation-drive-fails-uncover-black-money}} and nationwide increase in economic inequality did not decrease the popularity of `BJP' very significantly though a dip in popularity is observed in 2016 for all the news media houses.
For cloze test (c), we see Hindu and India Today both reporting similarly about the economy with higher negative words for economy in 2018 which resonates with the ground truth of GDP growth rate for India but ToI interestingly reports the opposite trend.

\section{Discussion}
\subsection{Is balance necessary?}

Many might argue that publicizable material begets news always and there is no point in considering balance despite being recommended by the experts of media watchdog groups\footnote{https://fair.org/about-fair/} \textsuperscript{,}\footnote{https://www.aim.org/about/who-we-are/} and the prestige press~\cite{lacy91}. In conformation with the viewpoint of the experts and the prestige press, we investigate the extent of imbalance here. In fact, this viewpoint is largely motivated by the study done by D’Alessio and Allen\cite{d2000media} since it is a standing evidence of such a balanced reporting environment. In a different country, in a different period which did not see the rise of social media and democratization of content creation, their studies confirm that balanced reporting (i.e., covering the newsmakers and the criticizers equally) is possible and was the norm for most of the newspapers in their specific setting.

\subsection{Bias: Then \& Now}
Lacy et al \cite{lacy91} showed that the prestige press was distinctly different from the other media outlets presenting a more balanced coverage of local stories compared to wide circulation media outlets.
D’Alessio and Allen~\cite{d2000media} in their meta-analysis considered 59 quantitative studies containing data concerned with partisan media bias in presidential election campaigns in the extended period of 1948--2000 but they found no significant bias in the newspapers or in the newsmagazines. Our study seems to contradict their findings(\cite{lacy91} is contradicted because all three newspapers in our study are highly read and respected for journalism and \cite{d2000media} is contradicted as they found little bias, we found significant bias) albeit in a completely different time window. Further while we studied the online news media, D’Alessio and Allen considered the print media in their study. We argue that the advent of Internet and its widespread availability has fundamentally changed the nature of news media over time. Hence the diffference is organic and provides validation to the claims reported by the media watchdogs or political parties\cite{f20}\cite{f21} \cite{f22} \cite{f23}  

The framing and the epistemological biases discussed by the authors in~\cite{recasens2013linguistic} also find relevance in our work. For instance, the framing bias corresponds to the tonality bias that we analyzed here. Similarly, the epistemological bias has parallels to the readability bias which is more subtle and harder to observe.

Finally, there is a significant difference in the way we perceive the notion of fairness. Many might argue that publicizable material begets news always and there is no point in considering equality of coverage as a measure of fairness despite that being recommended by the experts of media watchdog groups\cite{f6} and the prestige press~\cite{lacy91}. In conformation with this viewpoint of the experts and the prestige press, we develop the definition of the angular bias distance based on equality of importance. In fact, this viewpoint is also largely motivated by the study done by D’Alessio and Allen since it is a standing evidence of such a bias free reporting environment. In a different country, in a different period which did not see the rise of social media and democratization of content creation, their study confirms that bias-free reporting (i.e., covering the newsmakers and the criticizers equally) is possible and was the norm for most of the newspapers in their specific setting.

\subsection{Generalizability of the bias quantification framework}
In this section we discuss the generalizibility of the studies that we performed in the previous sections.

\noindent\textbf{Extending to other news media outlets}: Although we have used three news outlets for our analysis based on their online availability, the metrics that we propose are generic and can be easily computed for any other media outlet subject to data availability. Owing to the absence of digital archives we could not analyze some of the major players like Hindustan Times, The Economic Times, The Telegraph etc. However we are putting efforts to gather this data either by contacting the media houses directly or through appropriate digitization of the print version through the help of one of the national libraries. 

In fact since our metrics are very generic, this study can be easily extended to other countries subject to the online availability of English newspapers. We indeed plan to subset other countries from the Indian subcontinent including Sri Lanka, Bangladesh, Nepal and Burma that are socio-politically similar to India.

\noindent\textbf{Extending to a multi-lingual setting}: The eighth schedule to the Indian constitution lists as many as 22 scheduled languages. There are more than one daily newspapers published in each of these languages. Some of the major players\cite{f25} are \textit{Dainik Jagaran} (Hindi), \textit{Malayala Manorama} (Malayalam), \textit{Daily Thanthi} (Tamil), \textit{Lokmat} (Marathi) etc.  No analysis of media outlets in India is complete unless the study is done in a multi-lingual setting. Although our metrics are generic they are not language agnostic and would require processing multi-lingual text. However, NLP tools for every individual Indian language are not widely available. We plan to incorporate in future some of the languages where there are a few state-of-the-art NLP tools already available, e.g., Hindi, Bengali, Tamil etc.

\noindent\textbf{Extending to a multi-dimensional setting}: As also noted earlier we have restricted our study to only two major national parties (BJP and Congress). However, as per the latest reports from the Election Commission of India, there are 1841 registered parties. Eight of these are national parties, 52 are state parties while the rest are unrecognized parties~\cite{f26}. The bias vector therefore can theoretically have 1841 dimensions. However, in practice an interesting extension of the current study would be to at least factor in the eight national parties into the bias vector. Similarly, another important extension would be to study the state parties separately using a 52 dimension bias vector. However, the state parties would be typically active in the states, so this analysis would be more meaningful if performed on the state newspapers (English as well as regional languages). 

\noindent\textbf{Extending to a location specific setting}: All our analysis presented in the paper has been considering India as an individual geographic unit. However, we have already pointed out that there are differences in the number of articles mentioning different parts of India (location bias). One can therefore easily extend this study to factor in the location information present in the article. However, data for many of the locations would be extremely sparse; this study can possibly be done for some of the highly covered states only.

\subsection{In search for mitigation of bias}
 Our main objective in this paper was to introduce and quantify the different forms of bias that one is able to observe across the Indian news media outlets. In this section we shall try to outline some of the mechanisms that can be used to mitigate (at least partially) such biases.

\noindent\textbf{Making bias transparent}: This approach would envisage to make the user aware that he/she is consuming a biased news through various visualization techniques implemented on the online newspaper platforms making a topic-wise comparison between various media outlets. This might also include simple indicators like how much factual a news is or what part of the same news-story is a news item covering with what sentiment. Such a nudging practice is widely prevalent in the literature with objectives to deliver multiple aspects of news in social media~\cite{Park:2009} or for encouraging users to read about diverse political opinions~\cite{Munson:2010,Munson:2013}.

\noindent\textbf{Platform governance}: With the exponential penetration of the social media, the way users consume news has seen a sea of change. Most users active on different social media platforms now consume their daily news from the news stories that are recommended by these platforms\cite{f27}. With the explosion in the Indian smartphone market the number of such users is increasing in leaps and bounds. Such platforms can easily game the users to consume only biased political news~\cite{f28}. In February 2019 the United Kingdom's Digital, Media, Culture, and Sport (DCMS) committee issued a verdict in view of this rising problem. The verdict said that social media platforms can no longer hide themselves behind the claim that they are merely a `platform' and therefore have no responsibility of regulating the content they recommend to their users~\cite{f29}. In similar lines, the European Union has now issued the `EU Code of Conduct on Terror and Hate Content' (CoT) that applies to the entire EU region. EU, recently, has also deployed mechanisms to combat biased and fake news in the online world by constituting working groups that include voices from different avenues including academia, industry and civil society. For instance, in January 2018, 39 experts met to frame the `Code of Practice on Online Disinformation' which was signed by tech giants like Facebook, Google etc. We believe that Social System Researchers have a lead role to play in such committees and any code of conduct cannot materialize unless the effect of the algorithmic implementation of policies of these platforms are reexamined empirically.


\noindent\textbf{The road ahead.} Our main objective in this paper was to introduce and quantify the different forms of imbalance metrics that one is able to observe across the Indian news media outlets. This will subsequently help in better platform governance through informing readers about the extent of different kinds of imbalances present in an article they are consuming and by showing similar articles on same topic highlighting the opposite viewpoints in the recommended list helping to create a better inclusive worldview for the readers bursting the filter bubble of biased news consumption. In view of the recent resurgence of debates and legal actions around platform governance~\cite{f27,f29}, this, we believe, is a very significant step forward.

\if{0}
\subsection{Summary of observations} There are a few observations that recur all through our analysis. We narrate these below.\todo{@Souvic: Rewrite based on the new outline of this section.}
\begin{itemize}
    \item The trends in temporal variation of bias are very similar in case of every media house. 
    
    \item The overall profile structure of various imbalances show some similar trends. Despite the differences in the way each of these are measured, all of them point to a universal pattern thus reinforcing the existence and the pervasive nature of the pattern. In fact almost all of these profiles seem to be just shifted and scaled versions of each other (perhaps a combined effect of the influence from contemporary events and individual inclinations of the media houses). 
    \item The imbalance profiles of one media house maintain a consistent distance with other media houses for majority of the timeline or the whole of it for most of the measures showing the existence of consistent editorial level stylistic nuances.
    \item There is an occasional early or late change of trends for each newspaper in each metric relative to other media houses. These temporal shifts are visibly significant and consistent for significant portion (few months of shift for 3-5 years) over the observed timeline. However none of these seem to consistently span over the whole timeline. 
    \item An intriguing temporal trend seems to be persistent across most metrics. Across the imbalance measures involving coverage of content, positive sentiment, negative sentiment, subjectivity and use of superlatives \& comparatives, India Today shows early signs of change in trend relative to the other two news papers (early by several months in most of the cases).
    
\end{itemize}
\fi

\if{0}
\section{Correlation among linguistic factors}
It is possible that the linguistically motivated bias metrics that we studied in the previous section have some degree of correlation among themselves. Correlation analysis can give us at least two insights -- (i) if a particular pair of metric is highly correlated across all the news outlets then they should have some influence on one another and possibly define an equivalence class, (ii) if a particular pair of metric has high correlation for certain news outlets and low correlation for certain other then the differences can be attributed to the characteristics of the news outlets publishing the news. We present the Pearson's correlation coefficients between the different bias metrics in Figure~\ref{figpear} in cases where the observations are statistically significant ($p$-value $< 0.01$). First for all the news papers, most of the bias metrics are either anti-correlated or extremely loosely correlated among each other. For TOI, we observe an exception in one of the pairs as the sentiment imbalance and coverage imbalance seem to be somewhat correlated. To some extent this observation is also true for The Hindu. However, this is not true for India Today.
Possibly for TOI and The Hindu articles if there is a strong imbalance in coverage favoring a particular party then one should also be able to observe a similar strong sentiment imbalance favoring that party. 

\begin{figure*}[t]
\centering
\begin{subfigure}{0.65\columnwidth}
  \centering
  \includegraphics[trim=0 0 150 0, clip,width=\linewidth]{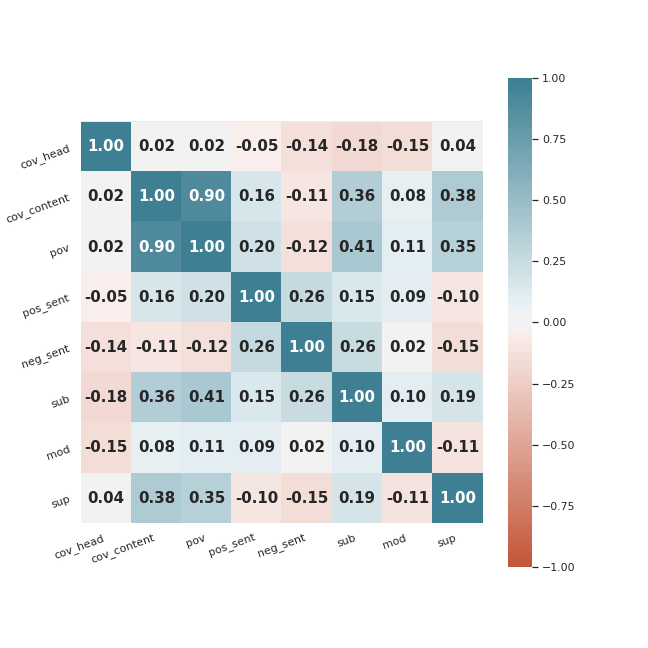}
  \caption{Times of India}
  \label{fig:sub1}
\end{subfigure}
\begin{subfigure}{0.65\columnwidth}
  \centering
  \includegraphics[trim=0 0 150 0, clip,width=\linewidth]{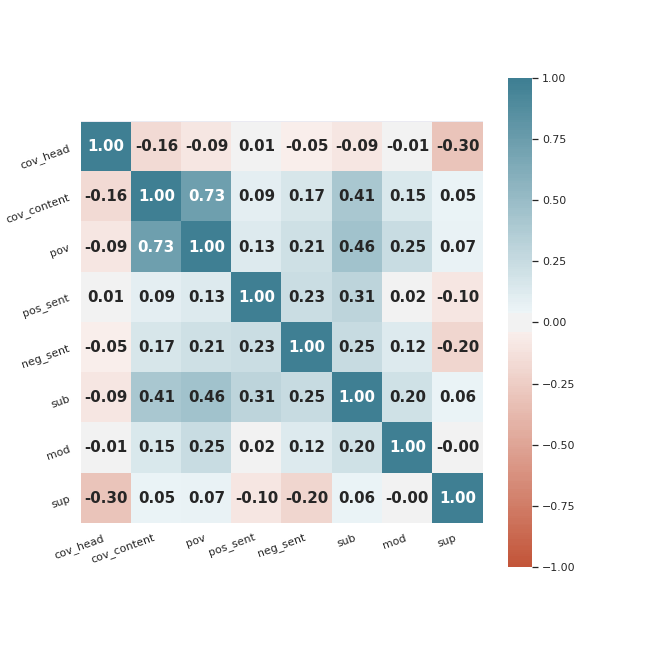}
  \caption{India Today}
  \label{fig:sub2}
\end{subfigure}
\begin{subfigure}{0.65\columnwidth}
  \centering
  \includegraphics[trim=0 0 50 0, clip,width=1.2\linewidth]{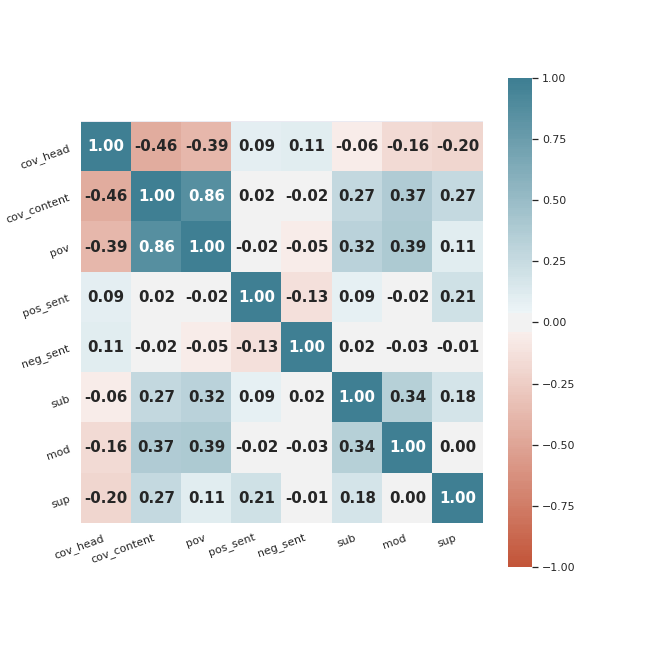}
  \caption{The Hindu}
  \label{fig:sub3}
\end{subfigure}%
\caption{\label{figpear}Spearman correlation coefficients between different imbalance metrics for the three newspapers. All observations are statistically significant ($p$-value $<0.01$). coverage: coverage imbalance, pov: point of view imbalance, possentiment: positive sentiment imbalance and negsentiment: negative sentiment imbalance.}
\end{figure*}
\fi

\if{0}
\subsection{What is the extent of religious polarization in India as depicted by these news media houses?}

Coming to the question of bias, we try to understand how religious polarization is portrayed by different news media houses. Hence, we employ similar methods like the previous section but in a discriminatory setting.

\noindent{\textbf{Coverage:}}
Before delving into the topic of religious polarization, we try to understand how religion is covered in political context in India by the newspapers in our corpus.
We plot the fraction of poltical articles where the names of any of the top seven religious identities in India (\textit{Hindu, Muslim, Sikh, Christian, Buddhist, Jain, Parsi}) has been covered by the newspapers over time in Figure~\ref{religion_fraction_fig}.

\begin{figure*}[t]
\centering
\begin{subfigure}{0.64\columnwidth}
  \centering
  \includegraphics[width=0.95\linewidth]{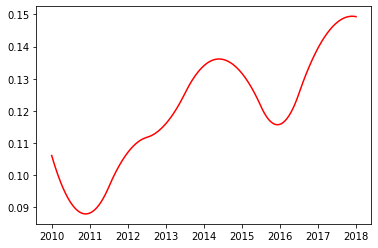}
  \caption{Times of India}
  \label{fig:sub1}
\end{subfigure}
\begin{subfigure}{0.64\columnwidth}
  \centering
  \includegraphics[width=0.95\linewidth]{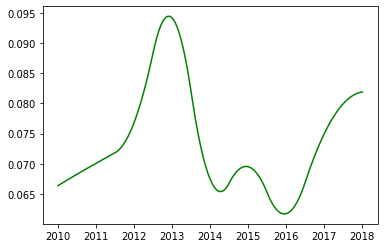}
  \caption{India Today}
  \label{fig:sub2}
\end{subfigure}
\begin{subfigure}{0.64\columnwidth}
  \centering 
  \includegraphics[width=0.95\linewidth]{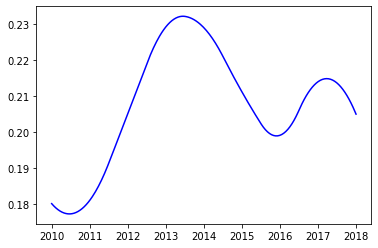}
  \caption{The Hindu}
  \label{fig:sub3}
\end{subfigure}%
\caption{\label{religion_fraction_fig}Fraction of articles using at least one of the names of the top 5 religions in India by population, plotted over the years.}
\end{figure*}

\begin{figure*}[t]
\centering
\begin{subfigure}{0.64\columnwidth}
  \centering
  \includegraphics[width=0.95\linewidth]{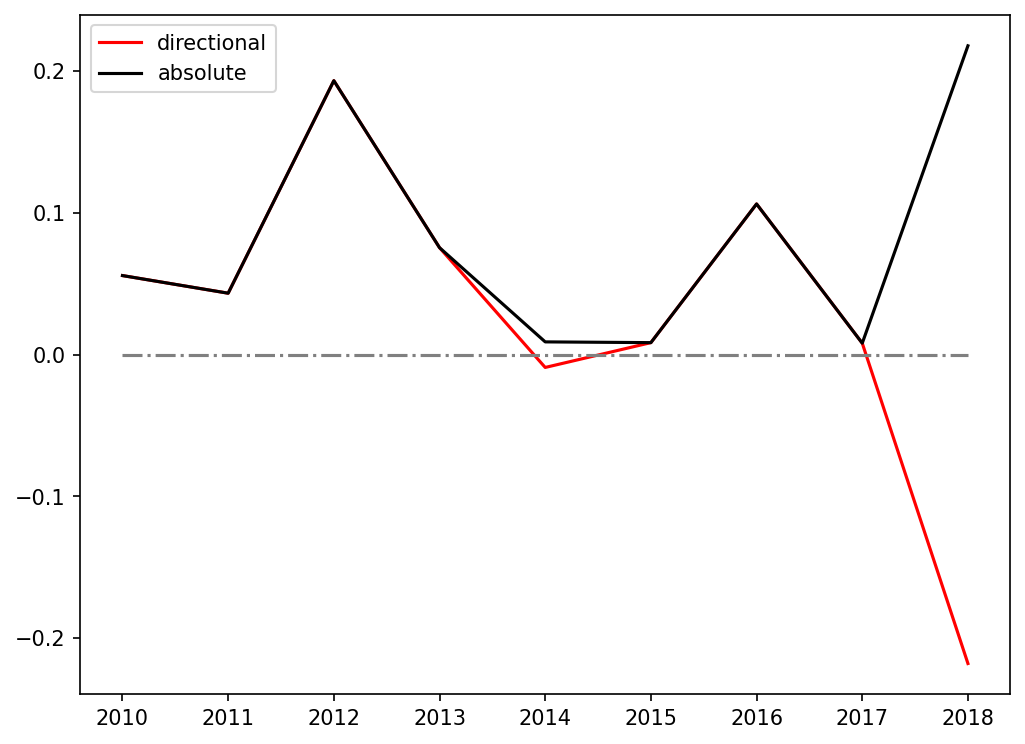}
  \caption{Times of India}
  \label{fig:sub1}
\end{subfigure}
\begin{subfigure}{0.64\columnwidth}
  \centering
  \includegraphics[width=0.95\linewidth]{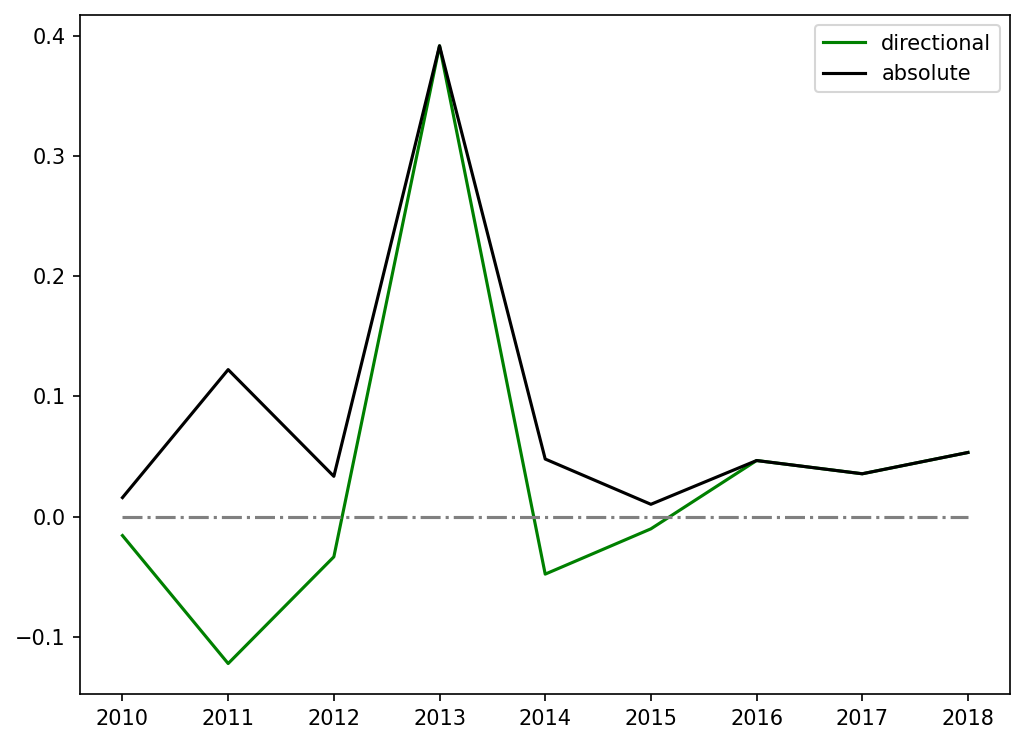}
  \caption{India Today}
  \label{fig:sub2}
\end{subfigure}
\begin{subfigure}{0.64\columnwidth}
  \centering
  \includegraphics[width=0.95\linewidth]{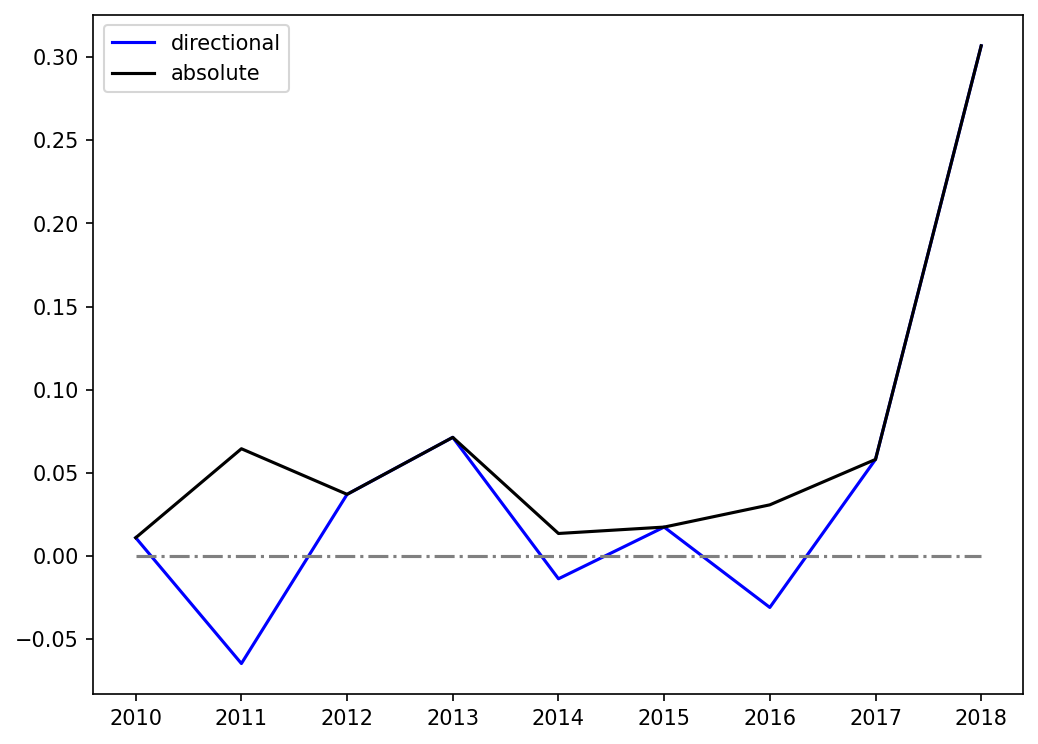}
  \caption{The Hindu}
  \label{fig:sub3}
\end{subfigure}%
\caption{\label{religious_polarization_fig}Extent of religious polarization quantified from the results of Cloze test 3. Direction for "directional" is affinity of Hindus towards BJP over vice versa as explained in the respective section.}
\end{figure*}

\noindent{\textbf{Polarization}}
Now we try to understand if the extent of polarization has also increased over time. We define polarization as the difference in voting preference of the two largest (and often in conflict) religious groups in India over the average voting preference.
Here we define voting preference of a group $\langle   x \rangle$  towards a specific political party $\langle   p \rangle$  as:

V($\langle   x \rangle$ ,$\langle   p \rangle$ ) = $P_{RoBERTa}$($\langle   mask \rangle$  = $\langle   p \rangle$  | input = ``$\langle   x \rangle$  will vote for $\langle   mask \rangle$ '') $\forall$ $\langle   x \rangle$  $\in$ \{``Hindus'', ``Muslims''\}, $\forall$ $\langle   p \rangle$  $\in$ \{``BJP'',``Congress''\}.

Further, affinity of the group is normalized to probabilities towards any of the two parties, arbitrarily selected to be \textit{BJP} (the mathematical formulation will only change in scale if the other party is chosen) as:
\begin{equation}
Pr(\langle   x \rangle )=\frac{V(\langle   x \rangle ,``BJP'')}{V(\langle   x \rangle ,``BJP'')+V(\langle   x \rangle ,``Congress'')}
\end{equation}
Now, polarization is defined as:
\begin{equation}
    Polarization (directionality\_towards\_Hindus) = \frac{Pr(``Hindus'')-Pr(``Muslims'')}{\frac{Pr(``Hindus'')+Pr(``Muslims'')}{2}}
\end{equation}

To get polarization value without directionality we ignore the sign of the value.
We calculated polarization for all the newspapers for each year using the 27 models trained and plotted in Figure~\ref{religious_polarization_fig}

\noindent{\textbf{Analysis:}}
In Figure~\ref{religion_fraction_fig}, we see the mentions of religions in political news is an increasing trend over the years 2010-2018 which is worrying as religion holds strong sentiment value and polarization on basis of religion can weaken a democracy. Also disturbingly, the mentions increase rapidly during poll time (2013-2014) and can be seen to be increasing before polling in 2019.

We see from Figure~\ref{religious_polarization_fig} that the pattern of change in religious polarization is different for each news media house. But, a consistent theme is that the extent of religious polarization is increasing over time from 2010 to 2018 for each newspaper. While for both the Hindu and India Today, Hindus and Muslims are shown to become polarized towards BJP and Congress respectively as is the case on the ground \textcolor{red}{refneeded}, Times of India shows the opposite trend, nevertheless increasing the extent of polarization over time. While we try to find the reason in detail from table \ref{relcloze_table}, we find that post 2014, all the news media groups show high popularity of BJP in India for both the groups. But popularity of BJP among muslims are portrayed to be higher in Times of India than other media groups and this gap only increases over the years. We find the results portrayed in the Hindu and India Today to be closer to reality than the situation consistently portrayed in Times of india\textcolor{red}{refneeded}. We see all three newsmedia groups show increasing polarization before both the national elections(pre-2014,pre-2018).
\fi

\if{0}
\subsection{How different are these newspapers from each other based on the Language model scores?}
We try to understand the characteristic difference of the three news media houses and how they evolved over time through the lens of RoBERTa.

MLM score( Masked Language Model score) is defined as:

$\sum_{s_j} \sum_i P(\langle   mask \rangle =token_i | s_{ji}) \times \log(P(\langle   mask \rangle =token_i | s_{ji}))$  where $s_{ji}$ is a sequence of tokens randomly taken from corpus j with the $i^{th}$ token masked.

In table.. we report the average MLM scores of the models trained and tested over different corpuses.
\noindent{\textbf{Analysis:}} To be written once the experiment is done. It is easy. It will take half an hour may be. We only report for 2010 and 2018.

\fi

\if{0}
\textcolor{red}{Population in each state vs coverage, statewise consumption vs coverage, online audience vs coverage pondering upon the reasons of difference, elections in states, cultural ignorance , language , homogenity etc.}
\textcolor{red}{change the figure's colorcode to red, yellow bright kind and maybe use loc2 bar graph figures}

\fi

\if{0}
\section{Can Media houses be analyzed like a person having personality traits?}

Personality traits~\cite{mathews2003} tend to reflect people's characteristic patterns of thoughts, feelings, and behaviors. They indicate how one could differ from  another in terms of where they stand on a set of basic trait dimensions that persist over time and across situations. One of the most popular ways to investigate these traits is the five-factor (big5) model~\cite{crae1992}. The model includes five primary dimensions (O-C-E-A-N). \textit{Openness} is the extent to which a person is open to experiencing different activities. \textit{Conscientiousness} is a person's tendency to act in an organized or thoughtful way. \textit{Extraversion} is a person's tendency to seek stimulation in the company of others. \textit{Agreeableness} is a person's tendency to be compassionate and cooperative toward others. \textit{Neuroticism} is the extent to which a person's emotions are sensitive to the person's environment.
 
 \begin{table}[ht]
\centering
\small
\begin{tabular}{c|c|c|c|c|c}
\hline
 & O& C& E& A& N \\ \hline
 
 ToI & 0.87 & 0.89 & 0.34& 0.22 & 0.08   \\ \hline 
 India Today & 0.90 & 0.88 & 0.39 & 0.26 & 0.07 \\ \hline
 Hindu & 0.81 & 0.90 & 0.29 & 0.28 & 0.06 \\ \hline

\end{tabular}%

\caption{\label{personality_traits_table} Five personality traits of the three newspapers aggregated over time (O: Openness, C: Conscientiousness, E: Extraversion, A: Agreeableness, N: Neuroticism). } 
\end{table}

\begin{figure*}[ht]
\centering
\begin{subfigure}{0.60\columnwidth}
  \centering
  \includegraphics[trim=85 0 180 0, clip,width=0.95\linewidth]{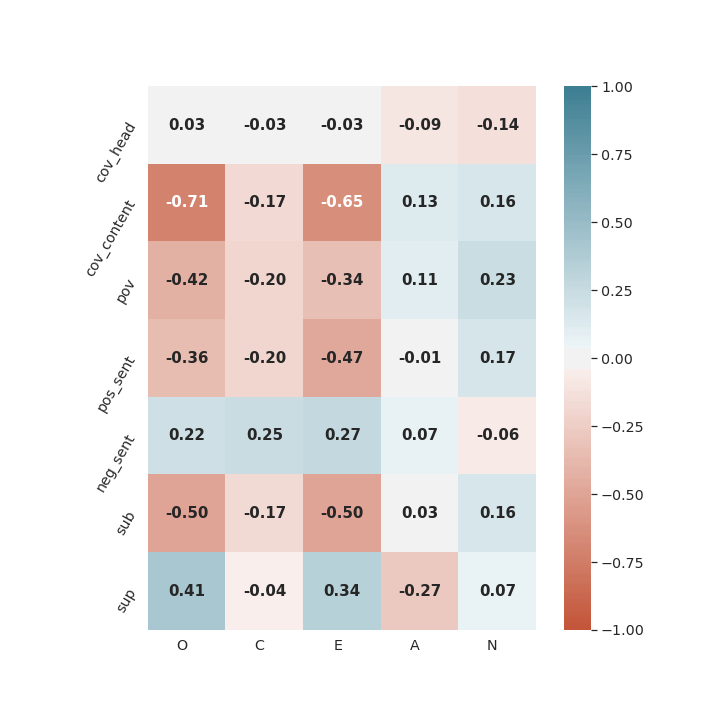}
  \caption{Times of India}
  \label{fig:sub1}
\end{subfigure}
\begin{subfigure}{0.60\columnwidth}
  \centering
  \includegraphics[trim=85 0 180 0, clip,width=0.95\linewidth]{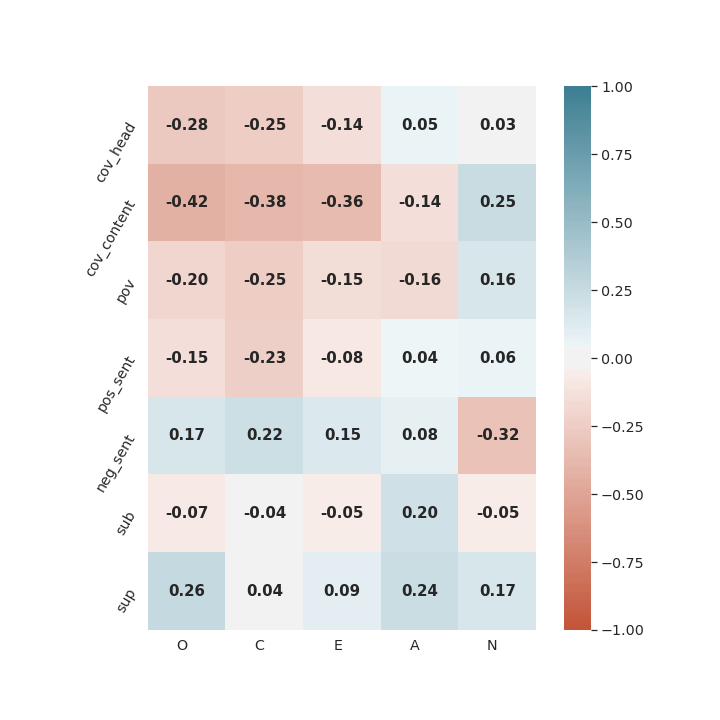}
  \caption{India Today}
  \label{fig:sub2}
\end{subfigure}
\begin{subfigure}{0.60\columnwidth}
  \centering
  \includegraphics[trim=85 0 75 0, clip,width=1.18\linewidth]{hindu_traits.png}
  \caption{The Hindu}
  \label{fig:sub3}
\end{subfigure}%
\caption{\label{personality_traits_fig}Pearson correlation coefficients between the big5 personality traits \& different imbalance metrics for the three newspapers. All observations are statistically significant ($p$-value $\langle   0.01$). coverage: coverage imbalance, pov: point of view imbalance, possentiment: positive sentiment imbalance and negsentiment: negative sentiment imbalance.}
\end{figure*}

 Significant research\cite{aaa,bbb} have been done to investigate the correlation of these personality traits with other more explicit internal characteristics like responsiveness to an event or specific reading preferences or participation in information sharing.
  Here we study how the personality traits reflected in the articles published from a media outlet are correlated to the different forms of previously defined imbalance scores. We compute the personality traits of a media outlet from all the article text published by that media outlet for each month and then average the values over time. We use the IBM Watson’s `Personality Insights' Tool\footnote{\url{https://cloud.ibm.com/docs/services/personality-insights/science.html}} to compute the personality traits.
 
 We summarize the main results in Table~\ref{personality_traits_table} and Figure~\ref{personality_traits_fig}. According to Table~\ref{personality_traits_table}, all three newspapers exhibit very similar traits since the journalistic vocabulary, from which these traits are computed, is largely shared. All the three media outlets rank high in openness and conscientiousness and very low on neuroticism. According to Figure \ref{personality_traits_fig}, openness is highly negatively correlated to the coverage imbalance for all the outlets very consistently. One way to read this observation is that high imbalance in coverage (in favour of BJP in this case) correlates well with lowered openness. BJP is found to advocate social and national conservatism quite openly and is often touted as a centre-right party. Congress, in contrast, is liberal on most issues and is touted as centre-left in ideological nature. So, by definition, ``openness'' is anti-correlated to ``conservatism'' and will be anti-correlated to a higher presence of BJP (conservative) related articles. A very similar observation also holds for the extraversion trait. This is especially true for TOI where higher positive sentiments, higher coverage of content, subjectivity or higher coverage of point of view for BJP correlates negatively with openness and extraversion. Overall reporting style of India Today shows little correlation with the personality traits. But The Hindu on the other hand, shows very unique characteristics.\fi 

\section{Conclusion \& future work}
In this paper we formulated and characterized imbalance in media through detailed analysis and discussion mostly in the context of political news. We empirically show the temporal relationship among the news outlets, the changing landscape of events featuring in them over time and the popularity trends of the political parties. 
\if{0}\begin{compactitem}
    \item We collected a huge dataset specific to the Indian news media with articles spanning over nine years for three leading newspapers to explore a more complex political environment than the US.
    \item We proposed linguistically motivated factors for quantification of imbalance. 
    \item We observed that there are certain universal trends in the imbalances across newspapers while there are some trends that are specific to a newspaper.
    \item We used advanced NLP tools to analyze and understand how India as depicted through the lens of these news media houses evolved through the last decade.
    \item We tried to understand if popularity of a particular topic over the other encourages more coverage of the topic.
    \item Finally, we analyze the personality traits reflected by the different media outlets and study their correlations with the different imbalance scores.
\end{compactitem}\fi

In future we would like to extend this work in multiple directions. One immediate task would be to see if we can study a wider range of media houses across different Indian languages and observe if we get similar results. Another immediate task would be to study evolution of religious and community-wise polarization in Indian society. Finally, we would also like to contribute to mitigation of such imbalances through platform governance. 

\bibliographystyle{IEEEtran}
\bibliography{ieee}

\newpage

\section{Biography Section}

\begin{IEEEbiographynophoto}{Souvic Chakraborty}
 is a PhD student in the Dept of Computer Science \& Engineering at Indian Institute of Technology, Kharagpur.
\end{IEEEbiographynophoto}
\begin{IEEEbiographynophoto}{Pawan Goyal}
 is an Associate Professor in the Dept of Computer Science \& Engineering at Indian Institute of Technology, Kharagpur.
\end{IEEEbiographynophoto}
\begin{IEEEbiographynophoto}{Animesh Mukherjee}
 is an Associate Professor (A K Singh Chair) in the Dept of Computer Science \& Engineering at Indian Institute of Technology, Kharagpur.
\end{IEEEbiographynophoto}

\vfill

\end{document}